\documentstyle[12pt]{article}
\input amssym.def
\input amssym.def
\input amssym
\def \dst {\displaystyle}
\def \txt {\textstyle}
\def \asabove{\leavevmode\vrule height 2pt depth -1.6pt width 23pt}
\def \ul{\underline}
\newcommand{\la} {\lambda}
\newcommand{\cs} {\check{S}}
\newcommand{\cu} {\check{u}}
\newcommand{\al} {\alpha}
\newcommand{\vet} {\vert \vec{\eta}\vert}
\newcommand{\be} {\beta}
\newcommand{\ga} {\gamma}
\newcommand{\bga}{\begin{array}{l}}
\newcommand{\ena}{\end{array}}
\newcommand{\bge}{\begin{equation}}
\newcommand{\ene}{\end{equation}}
\author{M. Zyskin \thanks{IHES, Le Bois-Marie 35, route de Chartres
F-91440 Bures-sur-Yvette, France; zyskin@math.ucr.edu} }
\title{A note on the glueball mass spectrum}

\date{June, 1998}

\begin{document}
\maketitle
\vspace{-200pt}
Preprint UCR 98-6.
\vspace{200pt}

\begin{abstract}
A conjectured duality between supergravity and $N=\infty$  gauge theories gives
predictions for the glueball masses as eigenvalues for a  supergravity
wave equations in a black hole geometry, and describes a physics, most close
toa high-temeperature expansion of a lattice QCD.
 We present an analytical solution
for eigenvalues and eigenfunctions, with eigenvalues given by zeroes of a
certain well-computable  function $r(p)$, the zeroes of which signifies that
two solutions with desired behaviour at
 two singular points become  linearly dependent.
Our computation shows corrections
to the WKB formula $ m^2= 6n(n+1)$  for eigenvalues corresponding to
 glueball masses in 3 dimensional QCD, and gives the first states with
masses $ m^2=$ 11.58766;  34.52698;  68.974962  ;\  114.91044;
172.33171;  241.236607;  321.626549, \ldots . In $QCD_4$, our
computation gives squares
of masses 37.169908;   81.354363;   138.473573;   208.859215;
 292.583628;  389.671368  500.132850,  623.97315 \ldots for
$O++$. In both cases, we have a powerful
method which allows to compute
eigenvalues with an arbitrary precision, if so  needed, 
which may provide quantative tests for the duality conjecture.  Our results
matches well with the numerical computation of \cite{oog} withing precision
reported there in both $QCD_3$ and $QCD_4$ cases. As an additional curiosity,
we report that for eigenvalues of about 7000, the power series, 
although convergent, has coefficients of  orders ${10}^{34}$; 
and also the final answer gets small, of  order ${10}^{-6}$ in $QCD_4$. 
Tricks were used to get reliably the function $r(p)$ above 7000. 
In principle we can go to infinitely high eigenavalues at an expence of 
computer sufferings; although eventually such computation will slow down to 
make it inpractical, also as corrections may be  expected for higher states, 
since  fine cancellations of very big terms is what produces higher  
eigenvalues. 

\end{abstract}
\newpage
\setcounter{section}{0}
\section{Glueball masses in $QCD_3$}.

The conjectures of Maldacena and Witten \cite{adsh}, \cite{adst}
gives, in particular,
 a prediction for glueball masses in $N=\infty$  QCD in  3 dimensions
from supergravity, namely, from certain solutions of classical equations of
motion in a black hole metric for a massless scalar dilaton field $\Phi$. 
The dilaton shows up  here  as it couples to the $O^{++}$ 
glueball operator; other glueballs, which couples to other fields, say a 
two-form in supergravity theory, are described, for example, in \cite{oog}. 
The  classical equations of motion for dilaton  are:
$$
\partial_\mu \left(\sqrt{g} g^{\mu \nu} \partial_\nu \Phi\right)
$$ 
in a black hole metric
$$
\dst\frac{ ds^2}{{l_s}^2 \sqrt{ 4 \pi g_s N}}= 
{\left(\rho^2 -\dst\frac{1}{\rho^2} \right)}^{-1} d\rho^2
+ \left(\rho^2 -\dst\frac{1}{\rho^2} \right) d\tau^2 + 
\rho^2 \sum_{i=1}^{3} {d x_i}^2 + {d \Omega_5}^{2}
$$
Here $x_1, x_2, x_3$ is where our $QCD_3$ lives, and it roughly should be
imagined as a 4-ball $x_1, x_2, x_3, \rho$ with $x_1, x_2, x_3$ on
$S^3$ boundary; $\tau$ is a circle, on which we pose antiperiodic
boundary conditions for fermions and periodic for bosons, thus 
breaking supersymmetry; $d\Omega_5$ is a sphere $S^5$ with a standart metric.
We  need to take solutions  of the form
$$
\Phi= f(\rho) e^{i k x}
$$
with the appropriate physical conditions for $\Phi$ at $\rho=\infty$, 
(normalizability), and at $\rho= 1$ (single-valuedness for complex $\rho$)
\cite{adst}. By introducing a variable $X=\rho^2$, (and calling it just
$x$ below) we get certain well-posed Sturm-Liouville problem:

\bge
\dst\frac{d^2 f}{dx^2} + \left(\dst\frac{1}{x} +
\dst\frac{1}{x-1}+\dst\frac{1}{x+1}\right)\dst\frac{df}{dx}
-\dst\frac{p}{x(x^2-1)} f =0 \label{glue3}
\ene
together with the boundary data:\\
\\
$f$ decays as $x\rightarrow \+\infty$, and: \\
$f$ is regular at $x=1$.
Here $4p = k^2$ is the mass of glueball states, $4p=-k^2$.

The above equation has regular singularities, and its solution can be written
as a series, with radius of convergence determined by distance between two
singularities. For the above equation, the radiuses of convergencies of
expansions at $1$ and infinity overlap, therefore, a reliable computation
is possible.

For an arbitrary $p$, the fundamental system of  solutions of the equation
(~\ref{glue3}) is the following:

For  $x$ with $1<x<\infty$, it is given by a linear combination of two
{\em convergent} series

\bge
\bga
y_1^{(\infty )}(x)= 1/(x^2)+  \dst\sum_{n=1}^{\infty } a[n] x^{-2-n}\\
y_2^{(\infty )}(x) = \frac{p^2}{2} Log (x)  y_1^{(\infty )}(x) +
\dst\sum_{n=1}^{\infty} b[n] x^{-n} \label{qcd3_inf}
\ena
\ene

where $a[n], b[n]$ are given by a recursion formula
$$
\bga
a[<0]=0;a[0]=1; a[n+1]=\dst\frac{1}{{(n+2)}^2-1}(a[n-1]({(n+1)}^2)+p a[n]);\\
\\
b[<0]=0;b[0]=0;\\
 b[n+1]= \dst\frac{1}{{(n)}^2-1}
\left( \frac{p^2}{2} (2 n a[n-1]-(2n-2)a[n-3])
+b[n-1] {(n-1)}^2 +p b[n]\right)
\ena
$$

For  $x$ with $0<x<2$, the fundamental system of  solutions is given by a
linear combination of two
{\em convergent} series

\bge
\bga
y_1^{(1)}(x)= 1+  \dst\sum_{n=1}^{\infty} a[n] {(x-1)}^{n}\\
y_2^{(1)}(x) = Log (x-1)  y_1^{(1)}(x) +
\dst\sum_{n=1}^{\infty} b[n] {(x-1)}^{n} \label{qcd3_1}
\ena
\ene

where $a[n], b[n]$ are given by a  recursion formula
$$
\bga
a[<0]=0;a[0]=1; a[n+1]=-\dst\frac{1}{2 {(n+1)}^2}
\left((3n(n+11)-p) a[n] + a[n-1]({n}^2-1)\right);\\
\\
b[<0]=0;b[0]=0;\\
b[n+1]=-\dst\frac{1}{2 {(n+1)}^2}
\left(4(n+1)a[n+1]+3(1+2n)a[n]+2n a[n-1]+
\right.
\\
+\left. (3n(n+1)-p) b[n] + b[n-1]({n}^2-1)\right)
\ena
$$

The solution of the boundary problem (\ref{glue3}) can be described as follows:
For certain $p$ there exist a solution which is proportional to $y_1^{(\infty
)}(x)$ for $x>1$ and to $y_1^{(1)}(x)$ as $|x-1|<1$. The condition for such a
solution to exist is that the Wronskian of two solutions $y_1^{(\infty )}(x)$
and  $y_1^{(1)}(x)$, which depend from $p$ as a parameter,
$$
Wronskain (p,x) =\pmatrix{y_1^{(1)}(x) & y_1^{(\infty )}(x)\cr
\frac{d}{dx} y_1^{(1)}(x) & \frac{d}{dx} y_1^{(\infty )}(x)}
$$
 must be zero. For $x$ such that $1< x< 2$ both the series defining
$y_1^{(\infty )}(x)$ and the series defining $y_1^{(1)}(x)$ are convergent,
therefore, the Wronskian can be computed. It is easy to see that the Wronskian
depend from $x$ as
$$
Wronskain (p,x) = \dst\frac{r(p)}{x(x-1)(x+1)},
$$
and therefore, the function $r(p)$
$$
r(p) = x(x-1)(x+1) Wronskain (p,x)
$$
can be computed at any point $x$ with $1<x<2$, which allows to determine
the function $r(p)$, with any desired accuracy, as the series 
(\ref{qcd3_inf}), (\ref{qcd3_1}) are convergent, and also the solutions
$y_1$ are analytic functions of $x$.

The spectrum of (~\ref{qlue3}) correspond to zeroes of the function $r(p)$.
By standart oscillation theorems, there are no zeroes of $r(p)$ for
$p\geq 0$, and there is an infinite discrete set of zeroes for $p<0$.
We  do not know an analytic formula for the roots of $r(p)$, however, 
we can find their numerical values, for example by plotting function $r(p)$
acurately and looking where the zeroes are, as one would do for a 
transcendental equation of the sort $\tan (x)=x$; we can also find roots
numerically with an arbitrary precision, for example by invoking Newton
method of finding a root. 
Such computations shows
that the first several roots are located at 
$ -11.58766;  -34.52698; - 68.974962  ;\  -114.91044; \ 
- 172.33171; \ - 241.236607;\   -321.626549, \ldots . $
We also believe that numbers  $4 p = - 6 n (n+1)$ , $n=1,2... $, which 
WKB method give,  are not the roots, although
they are close to those  roots listed.

\subsection{Orthogonality}
For the discrete set of values $p_n$ , $n=1,2\ldots $ such that 
$r(p)=0$ we have normalizable
at $1\leq x \leq\infty$ wave functions $\{ F_n (x)\}$ . 
Those functions  are orthogonal,
which here says
\bge
\dst\int_{1}^{+\infty} x (1-x^2) F_n (x) F_m (x) dx =0,\quad n\neq m
\ene

\subsection{Checking that -12 is not a root}
Since a WKB computation, which may or maynot have corrections in this case, 
shows the first eigenstate at $4p=-12$, we  were interested to check does our
function $r(p)$ has a root  exactly at $4p=-12$. It is easy to check whether 
or not it does, 
as  our series are  convergent (and pretty fast, faster then
the worst  of
of $\sum (-1)^n n (x-1)^n$ and $\sum (-1)^n n (1/x)^n$ do), for the $p$ we are
interested in). Therefore,
we can compute the function $r(p)$, using finite number of terms in the
series, and estimate the error. Thus, using our method, we can answer the
question whether or not the WKB formula gets corrections. We did such 
computation, and our result is 
that
$$ -0.022482 < {r(p)\vert}_{4p=-12} < -0.022481$$
(with the sign of $r(p)$ here consistent with our expectation to get the first
root close by with  $4p > -12 $, as $r(p)$ is positive and of order 1 for small
negative $p$). Thus we believe there is no root at exactly $ 4p=-12$, and
there is a correction to the WKB formula, with the exact root at about 
$-11.59$.

\section{Glueball spectrum in $QCD_4$}
The conjectures of Maldacena and Witten \cite{adsh}, \cite{adst}
gives, in particular,
 a prediction for glueball masses in $N=\infty$  QCD in  4 dimensions
from supergravity, namely, from certain solutions of classical equations of
motion in a black hole metric for a massless scalar dilaton field $\Phi$. 
The dilaton shows up  here  as it couples to the $O^{++}$ 
glueball operator; other glueballs, which couples to other fields, say a 
two-form in supergravity theory, are described, for example, in \cite{oog}. 
The  classical equations of motion for dilaton  are:
$$
\partial_\mu \left(\sqrt{g} g^{\mu \nu} \partial_\nu \Phi\right)
$$ 
where the metric is  a black hole
$$
\dst\frac{ ds^2}{{l_s}^2  {g_5}^2 N /4 \pi}= 
\dst\frac{d\rho^2}{4 \rho^{\frac{3}{2}}
\left(1 -\dst\frac{1}{\rho^3} \right)}
+ \rho^{\frac{3}{2}} \left(1 -\dst\frac{1}{\rho^3} \right) d\tau^2 + 
\rho^{\frac{3}{2}} \sum_{i=1}^{4} {d x_i}^2 + \rho^{\frac{1}{2}}
{d \Omega_4}^{2}
$$
Here $x_1, x_2, x_3, x_4 $ is where our $QCD_4$ lives, and, 
oversimplifing a bit,
it can be 
imagined as a 5-ball $x_1, x_2, x_3, x_4,  \rho$ with $x_1, x_2, x_3, x_4$ on
$S^4$ boundary; $\tau$ is a circle and $d\Omega_4$ is a sphere 
$S^4$ with a standart metric.
We  need to take solutions  of the form
$$
\Phi= F(\rho) e^{i k x}
$$
with the appropriate physical conditions for $\Phi$ at $\rho=\infty$, 
(normalizability), and at $\rho= 1$ (single-valuedness for complex $\rho$)
\cite{adst}. By introducing a variable $X=\rho^2$, (and calling it just
$x$ below) we get certain well-posed Sturm-Liouville problem:

\bge
(x^7-x) \dst\frac{d^2 F}{dx^2} + \left(10 x^6-4\right)\dst\frac{dF}{dx} -p x^3
F =0, \label{glue4}
\ene
(which is the equation of motion of a massles scalar field, dilaton, which
couples to the relevant glueball operator)\\
\vspace{5mm}

together with the boundary data:\\
\\
$F(x)$ decays as $x\rightarrow \+\infty$, and: \\
$F(x)$ is regular at $x=1$
Similarly to the equation (~\ref{glue3}), the singularities at $1$ and infinity
for this equation are regular, and the fundamental system of solutions
of the equation (~\ref{glue4}) can be written as power series with the
appropriate
radius of convergence, namely, for $1<x<\infty$ the fundamental system of
solutions is
is
\bge
\bga
y_1^{(\infty )}(x)= 1/(x^9)+  \dst\sum_{n=1}^{\infty } a[n] x^{-9-n}\\
y_2^{(\infty )}(x) = \frac{p^2}{2} Log (x)  y_1^{(\infty )}(x) +
\dst\sum_{n=1}^{\infty} b[n] x^{-n},
\ena
\ene

where $a[n]$  are given by a recursion formula
$$
a[<0]=0;a[0]=1; \\
$$
$$
a[2 n]= \dst\frac{1}{(2 n )(2n+9)}\left(p a[2n-2]+ (2 n )(2 n+3)
a[2n-6]\right);n=1,2,\ldots
$$
$$
$$

The series are convergent for $1<x<\infty$.

For  $x$ with $|x-1|\leq \dst\frac{\sqrt{3}-1}{\sqrt{2}}$, (with
$\dst\frac{\sqrt{3}-1}{\sqrt{2}}$ being  the distance between $1$ and the next
nearest
sixth roots  of one, $e^{\frac{\pm 2 \pi i}{6}}$, on a complex plane of $x$, as
can be discovered by looking
at the coefficient in front of the second derivative in our equation)
the fundamental system of  solutions is given by a linear combination of two
{\em convergent} series

\bge
\bga
y_1^{(1)}(x)= 1+  \dst\sum_{n=1}^{\infty} a[n] {(x-1)}^{n}\\
y_2^{(1)}(x) = Log (x-1)  y_1^{(1)}(x) +
\dst\sum_{n=1}^{\infty} b[n] {(x-1)}^{n},
\ena
\ene
where $a[n]$ are given by a  recursion formula
$$
a[<0]=0;a[0]=1;
$$
$$
\bga
a[n+1]=- \dst\frac{1}{6(n+1)^2}\left( ( n \ (21(n-1)+60)-p )a[n]+\right.
\\
\left. ((n-1)\ (35(n-2)+150)-3p)a[n-1]+ (\ (n-2)\ (35(n-3)+200)-3p)a[n-2]+
 \right.
\\
+\left. (\ (n-3)\ (21(n-4)+150)- p)  a[n-3]+((n-4)(7(n-5)+60))a[n-4]+ \right.
\\
+\left. ((n-5)(n-6+10))a[n-5] \right)
\ena
$$
Our boundary conditions require that the solution must be proportional
$y_1^{(\infty )}(x)$ for $x>1$ and to $y_1^{(1)}(x)$ as $|x-1|<1$, and
therefore,  the Wronskian of two solutions $y_1^{(\infty )}(x)$ and
$y_1^{(1)}(x)$,
$$
Wronskain (p,x) =\pmatrix{y_1^{(1)}(x) & y_1^{(\infty )}(x)\cr
\frac{d}{dx} y_1^{(1)}(x) & \frac{d}{dx} y_1^{(\infty )}(x)}
$$
which depend from $p$ as a parameter,
must be zero. For $x$ such that $1< x< \dst\frac{\sqrt{3}-1}{\sqrt{2}}$ the
series defining
$y_1^{(\infty )}(x)$ and the series defining $y_1^{(1)}(x)$ are both
convergent, (for any p), and therefore the Wronskian can be effectively
computed for such $x$ using our series. The Wronskian depend from $x$ as
follows:
$$
Wronskain (p,x) = \dst\frac{r(p)}{x^4 (x^6-1)}
$$
where
$$
r(p) =x^4 (x^6-1) Wronskain (p,x)
$$
does not depend from  $x$, and our eigenvalues are zeroes of the
function $r(p)$.

Since we can compute the Wronskian at any point 
 $1< x<\dst\frac{\sqrt{3}-1}{\sqrt{2}}$, using the series,  the function 
$r(p)$ is also
well- determined, and
we can examine where that function have zeroes, for examle by plotting
it and looking where it have zeroes, or, for a more precise computation,
invoking for example Newton method to find a root numerically.

\subsection{First few states}

The computation gives the first few zeroes at
-37.169908;\quad  -81.354363;\quad  -138.473573;\quad  -208.859215;
\quad -292.583628;\quad -389.671368\quad -500.1328  \quad -623.97315\ldots  .
There is an infinite discrete set of  roots, all roots are negative. It is
not difficult to get to roots for values of $p$ up to about $-4600$. 
Thereafter, computation is becoming increasingly difficult, as the coefficients
in the series grow very large, before being killed by powers of $x-1$ or
$1/x$. Also, the final product, our function $r(p)$, becomes pretty 
small, with coefficients of orders ${10}^{34}$ and results of orders
 ${10}^{-6}$, and we used tricks to go ahead. 

\subsection{Orthogonality}

For the discrete set of values $p_n$ , $n=1,2\ldots $ such that 
$r(p)=0$ we have normalizable
at $1\leq x \leq\infty$ wave functions $\{ F_n (x)\}$ . 
Those functions  are orthogonal,
which here says
\bge
\dst\int_{1}^{+\infty} x^4 (1-x^6) F_n (x) F_m (x) dx =0,\quad n\neq m
\ene

\section{Some pictures: $r(p)$ for $QCD_4$}
The graph of function $r(p)$ up to about $p=-7600$ for $QCD_4$ follows. The
roots of this function correspond to glueball eigenstate masses.
It is
not difficult to get to roots for values of $p$ up to about $-4600$. 
Thereafter, computation is becoming increasingly difficult, 
as for example for   p around 7000 there are terms of order 
${10}^{34}$ in the series, and although due to 
powers of 
of $x-1$ or
$1/x$ present the series is convergent, the first terms are very big, and in 
computing $r(p)$ they combine 
into an expresson of order ${10}^{-6}$; as brut-force computation failed here,
tricks were used to make it  
run for such (and
higher) eigenvalues. The eigenfunctions look pretty  weird here, and  it 
shouldn't take much for them and eigenvalues to be corrected by about 
anything; thus we suspect our hunt for higher eigenvalues is  for 
sport, mostly.
Surprisingly, function $r(p)$ keeps to be pretty nice there. ( Assistance of 
I.V.G. in putting up the pictures and fighting with latex is gratefully
acknowledged.)

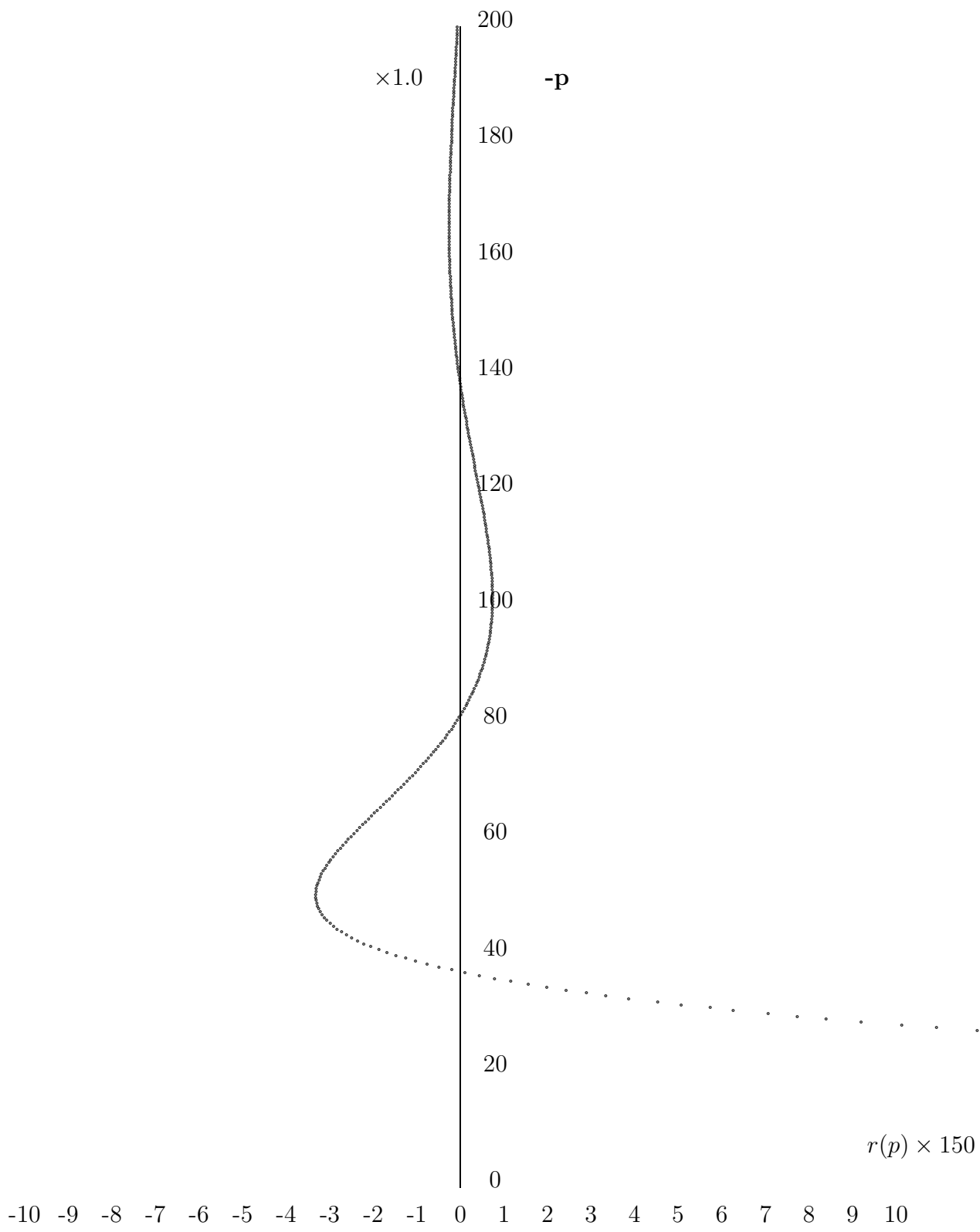
\setlength{\unitlength}{1mm}
\setcounter{enumi}{-11}
\setcounter{enumii}{-20}
\begin{picture}(150,200)
\put(75,0){\line(0,1){200}}
\multiput(0,-6)(7.5,0){21}{\addtocounter{enumi}{1}\makebox(0,0)[b]
{\arabic{enumi}}}
\multiput(81,0)(0,20){11}{\addtocounter{enumii}{20}\makebox(0,0)[b]
{\arabic{enumii}}}
\put(145,6){$r(p) \times 150$}
\put(88,190){\bf{ -p} }
\put(60,190){$\times 1.0 $}
\put(1330.,1.){\circle{0.4}}
\put(1280.,1.5){\circle{0.4}}
\put(1230.,2.){\circle{0.4}}
\put(1190.,2.5){\circle{0.4}}
\put(1150.,3.){\circle{0.4}}
\put(1110.,3.5){\circle{0.4}}
\put(1070.,4.){\circle{0.4}}
\put(1030.,4.5){\circle{0.4}}
\put(989.,5.){\circle{0.4}}
\put(952.,5.5){\circle{0.4}}
\put(917.,6.){\circle{0.4}}
\put(883.,6.5){\circle{0.4}}
\put(850.,7.){\circle{0.4}}
\put(818.,7.5){\circle{0.4}}
\put(787.,8.){\circle{0.4}}
\put(758.,8.5){\circle{0.4}}
\put(729.,9.){\circle{0.4}}
\put(701.,9.5){\circle{0.4}}
\put(674.,10.){\circle{0.4}}
\put(648.,10.5){\circle{0.4}}
\put(623.,11.){\circle{0.4}}
\put(599.,11.5){\circle{0.4}}
\put(575.,12.){\circle{0.4}}
\put(553.,12.5){\circle{0.4}}
\put(531.,13.){\circle{0.4}}
\put(510.,13.5){\circle{0.4}}
\put(490.,14.){\circle{0.4}}
\put(470.,14.5){\circle{0.4}}
\put(451.,15.){\circle{0.4}}
\put(433.,15.5){\circle{0.4}}
\put(416.,16.){\circle{0.4}}
\put(399.,16.5){\circle{0.4}}
\put(383.,17.){\circle{0.4}}
\put(367.,17.5){\circle{0.4}}
\put(352.,18.){\circle{0.4}}
\put(338.,18.5){\circle{0.4}}
\put(324.,19.){\circle{0.4}}
\put(310.,19.5){\circle{0.4}}
\put(298.,20.){\circle{0.4}}
\put(285.,20.5){\circle{0.4}}
\put(273.,21.){\circle{0.4}}
\put(262.,21.5){\circle{0.4}}
\put(251.,22.){\circle{0.4}}
\put(241.,22.5){\circle{0.4}}
\put(231.,23.){\circle{0.4}}
\put(221.,23.5){\circle{0.4}}
\put(212.,24.){\circle{0.4}}
\put(203.,24.5){\circle{0.4}}
\put(194.,25.){\circle{0.4}}
\put(186.,25.5){\circle{0.4}}
\put(178.,26.){\circle{0.4}}
\put(171.,26.5){\circle{0.4}}
\put(164.,27.){\circle{0.4}}
\put(157.,27.5){\circle{0.4}}
\put(151.,28.){\circle{0.4}}
\put(144.,28.5){\circle{0.4}}
\put(138.,29.){\circle{0.4}}
\put(133.,29.5){\circle{0.4}}
\put(128.,30.){\circle{0.4}}
\put(122.,30.5){\circle{0.4}}
\put(118.,31.){\circle{0.4}}
\put(113.,31.5){\circle{0.4}}
\put(109.,32.){\circle{0.4}}
\put(104.,32.5){\circle{0.4}}
\put(100.,33.){\circle{0.4}}
\put(96.7,33.5){\circle{0.4}}
\put(93.2,34.){\circle{0.4}}
\put(89.9,34.5){\circle{0.4}}
\put(86.7,35.){\circle{0.4}}
\put(83.7,35.5){\circle{0.4}}
\put(80.9,36.){\circle{0.4}}
\put(78.3,36.5){\circle{0.4}}
\put(75.8,37.){\circle{0.4}}
\put(73.5,37.5){\circle{0.4}}
\put(71.3,38.){\circle{0.4}}
\put(69.3,38.5){\circle{0.4}}
\put(67.3,39.){\circle{0.4}}
\put(65.6,39.5){\circle{0.4}}
\put(63.9,40.){\circle{0.4}}
\put(62.4,40.5){\circle{0.4}}
\put(61.,41.){\circle{0.4}}
\put(59.6,41.5){\circle{0.4}}
\put(58.4,42.){\circle{0.4}}
\put(57.3,42.5){\circle{0.4}}
\put(56.3,43.){\circle{0.4}}
\put(55.4,43.5){\circle{0.4}}
\put(54.6,44.){\circle{0.4}}
\put(53.8,44.5){\circle{0.4}}
\put(53.2,45.){\circle{0.4}}
\put(52.6,45.5){\circle{0.4}}
\put(52.,46.){\circle{0.4}}
\put(51.6,46.5){\circle{0.4}}
\put(51.2,47.){\circle{0.4}}
\put(50.9,47.5){\circle{0.4}}
\put(50.6,48.){\circle{0.4}}
\put(50.4,48.5){\circle{0.4}}
\put(50.3,49.){\circle{0.4}}
\put(50.2,49.5){\circle{0.4}}
\put(50.1,50.){\circle{0.4}}
\put(50.1,50.5){\circle{0.4}}
\put(50.2,51.){\circle{0.4}}
\put(50.2,51.5){\circle{0.4}}
\put(50.3,52.){\circle{0.4}}
\put(50.5,52.5){\circle{0.4}}
\put(50.7,53.){\circle{0.4}}
\put(50.9,53.5){\circle{0.4}}
\put(51.1,54.){\circle{0.4}}
\put(51.4,54.5){\circle{0.4}}
\put(51.7,55.){\circle{0.4}}
\put(52.,55.5){\circle{0.4}}
\put(52.4,56.){\circle{0.4}}
\put(52.7,56.5){\circle{0.4}}
\put(53.1,57.){\circle{0.4}}
\put(53.5,57.5){\circle{0.4}}
\put(53.9,58.){\circle{0.4}}
\put(54.4,58.5){\circle{0.4}}
\put(54.8,59.){\circle{0.4}}
\put(55.3,59.5){\circle{0.4}}
\put(55.7,60.){\circle{0.4}}
\put(56.2,60.5){\circle{0.4}}
\put(56.7,61.){\circle{0.4}}
\put(57.2,61.5){\circle{0.4}}
\put(57.7,62.){\circle{0.4}}
\put(58.2,62.5){\circle{0.4}}
\put(58.7,63.){\circle{0.4}}
\put(59.2,63.5){\circle{0.4}}
\put(59.7,64.){\circle{0.4}}
\put(60.2,64.5){\circle{0.4}}
\put(60.7,65.){\circle{0.4}}
\put(61.3,65.5){\circle{0.4}}
\put(61.8,66.){\circle{0.4}}
\put(62.3,66.5){\circle{0.4}}
\put(62.8,67.){\circle{0.4}}
\put(63.3,67.5){\circle{0.4}}
\put(63.8,68.){\circle{0.4}}
\put(64.3,68.5){\circle{0.4}}
\put(64.8,69.){\circle{0.4}}
\put(65.3,69.5){\circle{0.4}}
\put(65.8,70.){\circle{0.4}}
\put(66.3,70.5){\circle{0.4}}
\put(66.8,71.){\circle{0.4}}
\put(67.3,71.5){\circle{0.4}}
\put(67.7,72.){\circle{0.4}}
\put(68.2,72.5){\circle{0.4}}
\put(68.6,73.){\circle{0.4}}
\put(69.1,73.5){\circle{0.4}}
\put(69.5,74.){\circle{0.4}}
\put(70.,74.5){\circle{0.4}}
\put(70.4,75.){\circle{0.4}}
\put(70.8,75.5){\circle{0.4}}
\put(71.2,76.){\circle{0.4}}
\put(71.6,76.5){\circle{0.4}}
\put(72.,77.){\circle{0.4}}
\put(72.4,77.5){\circle{0.4}}
\put(72.7,78.){\circle{0.4}}
\put(73.1,78.5){\circle{0.4}}
\put(73.5,79.){\circle{0.4}}
\put(73.8,79.5){\circle{0.4}}
\put(74.1,80.){\circle{0.4}}
\put(74.5,80.5){\circle{0.4}}
\put(74.8,81.){\circle{0.4}}
\put(75.1,81.5){\circle{0.4}}
\put(75.4,82.){\circle{0.4}}
\put(75.7,82.5){\circle{0.4}}
\put(76.,83.){\circle{0.4}}
\put(76.2,83.5){\circle{0.4}}
\put(76.5,84.){\circle{0.4}}
\put(76.7,84.5){\circle{0.4}}
\put(77.,85.){\circle{0.4}}
\put(77.2,85.5){\circle{0.4}}
\put(77.4,86.){\circle{0.4}}
\put(77.7,86.5){\circle{0.4}}
\put(77.9,87.){\circle{0.4}}
\put(78.1,87.5){\circle{0.4}}
\put(78.3,88.){\circle{0.4}}
\put(78.4,88.5){\circle{0.4}}
\put(78.6,89.){\circle{0.4}}
\put(78.8,89.5){\circle{0.4}}
\put(78.9,90.){\circle{0.4}}
\put(79.1,90.5){\circle{0.4}}
\put(79.2,91.){\circle{0.4}}
\put(79.4,91.5){\circle{0.4}}
\put(79.5,92.){\circle{0.4}}
\put(79.6,92.5){\circle{0.4}}
\put(79.7,93.){\circle{0.4}}
\put(79.8,93.5){\circle{0.4}}
\put(79.9,94.){\circle{0.4}}
\put(80.,94.5){\circle{0.4}}
\put(80.1,95.){\circle{0.4}}
\put(80.2,95.5){\circle{0.4}}
\put(80.2,96.){\circle{0.4}}
\put(80.3,96.5){\circle{0.4}}
\put(80.3,97.){\circle{0.4}}
\put(80.4,97.5){\circle{0.4}}
\put(80.4,98.){\circle{0.4}}
\put(80.5,98.5){\circle{0.4}}
\put(80.5,99.){\circle{0.4}}
\put(80.5,99.5){\circle{0.4}}
\put(80.5,100.){\circle{0.4}}
\put(80.6,100.5){\circle{0.4}}
\put(80.6,101.){\circle{0.4}}
\put(80.6,101.5){\circle{0.4}}
\put(80.6,102.){\circle{0.4}}
\put(80.6,102.5){\circle{0.4}}
\put(80.5,103.){\circle{0.4}}
\put(80.5,103.5){\circle{0.4}}
\put(80.5,104.){\circle{0.4}}
\put(80.5,104.5){\circle{0.4}}
\put(80.5,105.){\circle{0.4}}
\put(80.4,105.5){\circle{0.4}}
\put(80.4,106.){\circle{0.4}}
\put(80.3,106.5){\circle{0.4}}
\put(80.3,107.){\circle{0.4}}
\put(80.3,107.5){\circle{0.4}}
\put(80.2,108.){\circle{0.4}}
\put(80.2,108.5){\circle{0.4}}
\put(80.1,109.){\circle{0.4}}
\put(80.,109.5){\circle{0.4}}
\put(80.,110.){\circle{0.4}}
\put(79.9,110.5){\circle{0.4}}
\put(79.8,111.){\circle{0.4}}
\put(79.8,111.5){\circle{0.4}}
\put(79.7,112.){\circle{0.4}}
\put(79.6,112.5){\circle{0.4}}
\put(79.5,113.){\circle{0.4}}
\put(79.5,113.5){\circle{0.4}}
\put(79.4,114.){\circle{0.4}}
\put(79.3,114.5){\circle{0.4}}
\put(79.2,115.){\circle{0.4}}
\put(79.1,115.5){\circle{0.4}}
\put(79.1,116.){\circle{0.4}}
\put(79.,116.5){\circle{0.4}}
\put(78.9,117.){\circle{0.4}}
\put(78.8,117.5){\circle{0.4}}
\put(78.7,118.){\circle{0.4}}
\put(78.6,118.5){\circle{0.4}}
\put(78.5,119.){\circle{0.4}}
\put(78.4,119.5){\circle{0.4}}
\put(78.3,120.){\circle{0.4}}
\put(78.2,120.5){\circle{0.4}}
\put(78.1,121.){\circle{0.4}}
\put(78.,121.5){\circle{0.4}}
\put(77.9,122.){\circle{0.4}}
\put(77.8,122.5){\circle{0.4}}
\put(77.7,123.){\circle{0.4}}
\put(77.6,123.5){\circle{0.4}}
\put(77.5,124.){\circle{0.4}}
\put(77.5,124.5){\circle{0.4}}
\put(77.4,125.){\circle{0.4}}
\put(77.3,125.5){\circle{0.4}}
\put(77.2,126.){\circle{0.4}}
\put(77.1,126.5){\circle{0.4}}
\put(77.,127.){\circle{0.4}}
\put(76.9,127.5){\circle{0.4}}
\put(76.8,128.){\circle{0.4}}
\put(76.7,128.5){\circle{0.4}}
\put(76.6,129.){\circle{0.4}}
\put(76.5,129.5){\circle{0.4}}
\put(76.4,130.){\circle{0.4}}
\put(76.3,130.5){\circle{0.4}}
\put(76.2,131.){\circle{0.4}}
\put(76.1,131.5){\circle{0.4}}
\put(76.1,132.){\circle{0.4}}
\put(76.,132.5){\circle{0.4}}
\put(75.9,133.){\circle{0.4}}
\put(75.8,133.5){\circle{0.4}}
\put(75.7,134.){\circle{0.4}}
\put(75.6,134.5){\circle{0.4}}
\put(75.5,135.){\circle{0.4}}
\put(75.5,135.5){\circle{0.4}}
\put(75.4,136.){\circle{0.4}}
\put(75.3,136.5){\circle{0.4}}
\put(75.2,137.){\circle{0.4}}
\put(75.1,137.5){\circle{0.4}}
\put(75.1,138.){\circle{0.4}}
\put(75.,138.5){\circle{0.4}}
\put(74.9,139.){\circle{0.4}}
\put(74.9,139.5){\circle{0.4}}
\put(74.8,140.){\circle{0.4}}
\put(74.7,140.5){\circle{0.4}}
\put(74.6,141.){\circle{0.4}}
\put(74.6,141.5){\circle{0.4}}
\put(74.5,142.){\circle{0.4}}
\put(74.5,142.5){\circle{0.4}}
\put(74.4,143.){\circle{0.4}}
\put(74.3,143.5){\circle{0.4}}
\put(74.3,144.){\circle{0.4}}
\put(74.2,144.5){\circle{0.4}}
\put(74.2,145.){\circle{0.4}}
\put(74.1,145.5){\circle{0.4}}
\put(74.1,146.){\circle{0.4}}
\put(74.,146.5){\circle{0.4}}
\put(74.,147.){\circle{0.4}}
\put(73.9,147.5){\circle{0.4}}
\put(73.9,148.){\circle{0.4}}
\put(73.8,148.5){\circle{0.4}}
\put(73.8,149.){\circle{0.4}}
\put(73.7,149.5){\circle{0.4}}
\put(73.7,150.){\circle{0.4}}
\put(73.6,150.5){\circle{0.4}}
\put(73.6,151.){\circle{0.4}}
\put(73.6,151.5){\circle{0.4}}
\put(73.5,152.){\circle{0.4}}
\put(73.5,152.5){\circle{0.4}}
\put(73.5,153.){\circle{0.4}}
\put(73.4,153.5){\circle{0.4}}
\put(73.4,154.){\circle{0.4}}
\put(73.4,154.5){\circle{0.4}}
\put(73.4,155.){\circle{0.4}}
\put(73.3,155.5){\circle{0.4}}
\put(73.3,156.){\circle{0.4}}
\put(73.3,156.5){\circle{0.4}}
\put(73.3,157.){\circle{0.4}}
\put(73.2,157.5){\circle{0.4}}
\put(73.2,158.){\circle{0.4}}
\put(73.2,158.5){\circle{0.4}}
\put(73.2,159.){\circle{0.4}}
\put(73.2,159.5){\circle{0.4}}
\put(73.2,160.){\circle{0.4}}
\put(73.1,160.5){\circle{0.4}}
\put(73.1,161.){\circle{0.4}}
\put(73.1,161.5){\circle{0.4}}
\put(73.1,162.){\circle{0.4}}
\put(73.1,162.5){\circle{0.4}}
\put(73.1,163.){\circle{0.4}}
\put(73.1,163.5){\circle{0.4}}
\put(73.1,164.){\circle{0.4}}
\put(73.1,164.5){\circle{0.4}}
\put(73.1,165.){\circle{0.4}}
\put(73.1,165.5){\circle{0.4}}
\put(73.1,166.){\circle{0.4}}
\put(73.1,166.5){\circle{0.4}}
\put(73.1,167.){\circle{0.4}}
\put(73.1,167.5){\circle{0.4}}
\put(73.1,168.){\circle{0.4}}
\put(73.1,168.5){\circle{0.4}}
\put(73.1,169.){\circle{0.4}}
\put(73.1,169.5){\circle{0.4}}
\put(73.1,170.){\circle{0.4}}
\put(73.1,170.5){\circle{0.4}}
\put(73.1,171.){\circle{0.4}}
\put(73.2,171.5){\circle{0.4}}
\put(73.2,172.){\circle{0.4}}
\put(73.2,172.5){\circle{0.4}}
\put(73.2,173.){\circle{0.4}}
\put(73.2,173.5){\circle{0.4}}
\put(73.2,174.){\circle{0.4}}
\put(73.2,174.5){\circle{0.4}}
\put(73.3,175.){\circle{0.4}}
\put(73.3,175.5){\circle{0.4}}
\put(73.3,176.){\circle{0.4}}
\put(73.3,176.5){\circle{0.4}}
\put(73.3,177.){\circle{0.4}}
\put(73.3,177.5){\circle{0.4}}
\put(73.4,178.){\circle{0.4}}
\put(73.4,178.5){\circle{0.4}}
\put(73.4,179.){\circle{0.4}}
\put(73.4,179.5){\circle{0.4}}
\put(73.5,180.){\circle{0.4}}
\put(73.5,180.5){\circle{0.4}}
\put(73.5,181.){\circle{0.4}}
\put(73.5,181.5){\circle{0.4}}
\put(73.5,182.){\circle{0.4}}
\put(73.6,182.5){\circle{0.4}}
\put(73.6,183.){\circle{0.4}}
\put(73.6,183.5){\circle{0.4}}
\put(73.6,184.){\circle{0.4}}
\put(73.7,184.5){\circle{0.4}}
\put(73.7,185.){\circle{0.4}}
\put(73.7,185.5){\circle{0.4}}
\put(73.7,186.){\circle{0.4}}
\put(73.8,186.5){\circle{0.4}}
\put(73.8,187.){\circle{0.4}}
\put(73.8,187.5){\circle{0.4}}
\put(73.9,188.){\circle{0.4}}
\put(73.9,188.5){\circle{0.4}}
\put(73.9,189.){\circle{0.4}}
\put(73.9,189.5){\circle{0.4}}
\put(74.,190.){\circle{0.4}}
\put(74.,190.5){\circle{0.4}}
\put(74.,191.){\circle{0.4}}
\put(74.1,191.5){\circle{0.4}}
\put(74.1,192.){\circle{0.4}}
\put(74.1,192.5){\circle{0.4}}
\put(74.1,193.){\circle{0.4}}
\put(74.2,193.5){\circle{0.4}}
\put(74.2,194.){\circle{0.4}}
\put(74.2,194.5){\circle{0.4}}
\put(74.3,195.){\circle{0.4}}
\put(74.3,195.5){\circle{0.4}}
\put(74.3,196.){\circle{0.4}}
\put(74.3,196.5){\circle{0.4}}
\put(74.4,197.){\circle{0.4}}
\put(74.4,197.5){\circle{0.4}}
\put(74.4,198.){\circle{0.4}}
\put(74.5,198.5){\circle{0.4}}
\put(74.5,199.){\circle{0.4}}
\put(74.5,199.5){\circle{0.4}}
\put(74.5,200.){\circle{0.4}}

\end{picture}

\newpage
\setlength{\unitlength}{1mm}
\setcounter{enumi}{-11}
\setcounter{enumii}{110}
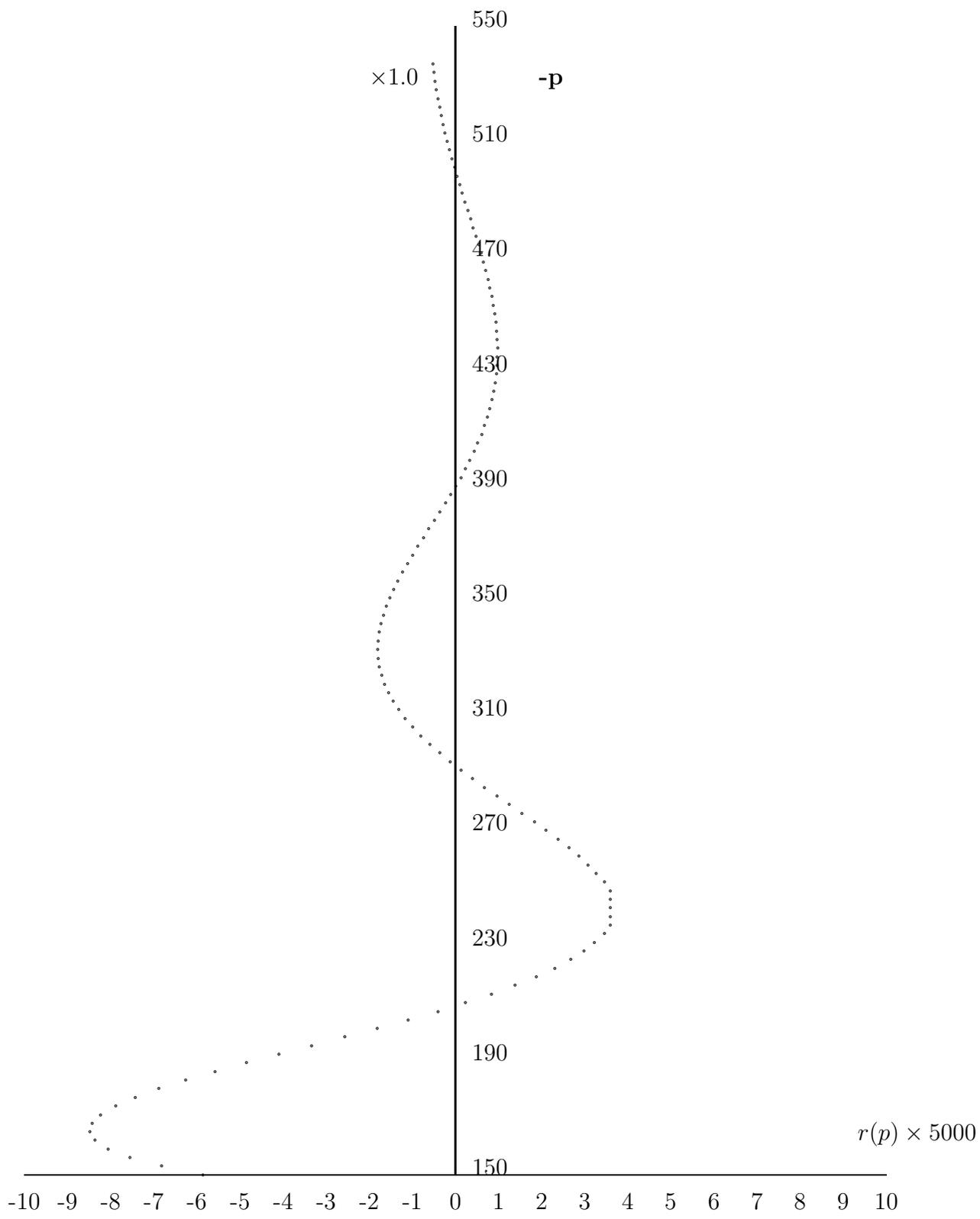
\begin{picture}(150,200)
\put(75,0){\line(0,1){200}}
\multiput(0,-6)(7.5,0){21}{\addtocounter{enumi}{1}\makebox(0,0)[b]
{\arabic{enumi}}}
\multiput(81,0)(0,20){11}{\addtocounter{enumii}{40}\makebox(0,0)[b]
{\arabic{enumii}}}
\put(145,6){$r(p) \times 5000$}
\put(88,190){\bf{ -p} }
\put(60,190){$\times 1.0 $}
\put(0,0){\line(1,0){150}}
\put(31.1,0){\circle{0.4}}
\put(23.9,1.5){\circle{0.4}}
\put(18.4,3.){\circle{0.4}}
\put(14.6,4.5){\circle{0.4}}
\put(12.3,6.){\circle{0.4}}
\put(11.4,7.5){\circle{0.4}}
\put(11.8,9.){\circle{0.4}}
\put(13.3,10.5){\circle{0.4}}
\put(15.9,12.){\circle{0.4}}
\put(19.3,13.5){\circle{0.4}}
\put(23.4,15.){\circle{0.4}}
\put(28.1,16.5){\circle{0.4}}
\put(33.2,18.){\circle{0.4}}
\put(38.7,19.5){\circle{0.4}}
\put(44.3,21.){\circle{0.4}}
\put(50.,22.5){\circle{0.4}}
\put(55.8,24.){\circle{0.4}}
\put(61.4,25.5){\circle{0.4}}
\put(66.8,27.){\circle{0.4}}
\put(72.,28.5){\circle{0.4}}
\put(76.8,30.){\circle{0.4}}
\put(81.3,31.5){\circle{0.4}}
\put(85.4,33.){\circle{0.4}}
\put(89.1,34.5){\circle{0.4}}
\put(92.3,36.){\circle{0.4}}
\put(95.1,37.5){\circle{0.4}}
\put(97.4,39.){\circle{0.4}}
\put(99.2,40.5){\circle{0.4}}
\put(101.,42.){\circle{0.4}}
\put(102.,43.5){\circle{0.4}}
\put(102.,45.){\circle{0.4}}
\put(102.,46.5){\circle{0.4}}
\put(102.,48.){\circle{0.4}}
\put(102.,49.5){\circle{0.4}}
\put(101.,51.){\circle{0.4}}
\put(99.5,52.5){\circle{0.4}}
\put(98.1,54.){\circle{0.4}}
\put(96.5,55.5){\circle{0.4}}
\put(94.8,57.){\circle{0.4}}
\put(92.9,58.5){\circle{0.4}}
\put(90.8,60.){\circle{0.4}}
\put(88.8,61.5){\circle{0.4}}
\put(86.6,63.){\circle{0.4}}
\put(84.4,64.5){\circle{0.4}}
\put(82.3,66.){\circle{0.4}}
\put(80.1,67.5){\circle{0.4}}
\put(78.,69.){\circle{0.4}}
\put(76.,70.5){\circle{0.4}}
\put(74.1,72.){\circle{0.4}}
\put(72.3,73.5){\circle{0.4}}
\put(70.6,75.){\circle{0.4}}
\put(69.,76.5){\circle{0.4}}
\put(67.6,78.){\circle{0.4}}
\put(66.3,79.5){\circle{0.4}}
\put(65.2,81.){\circle{0.4}}
\put(64.2,82.5){\circle{0.4}}
\put(63.4,84.){\circle{0.4}}
\put(62.7,85.5){\circle{0.4}}
\put(62.2,87.){\circle{0.4}}
\put(61.8,88.5){\circle{0.4}}
\put(61.6,90.){\circle{0.4}}
\put(61.5,91.5){\circle{0.4}}
\put(61.6,93.){\circle{0.4}}
\put(61.8,94.5){\circle{0.4}}
\put(62.1,96.){\circle{0.4}}
\put(62.5,97.5){\circle{0.4}}
\put(63.,99.){\circle{0.4}}
\put(63.6,100.5){\circle{0.4}}
\put(64.3,102.){\circle{0.4}}
\put(65.,103.5){\circle{0.4}}
\put(65.8,105.){\circle{0.4}}
\put(66.7,106.5){\circle{0.4}}
\put(67.6,108.){\circle{0.4}}
\put(68.5,109.5){\circle{0.4}}
\put(69.5,111.){\circle{0.4}}
\put(70.4,112.5){\circle{0.4}}
\put(71.4,114.){\circle{0.4}}
\put(72.3,115.5){\circle{0.4}}
\put(73.3,117.){\circle{0.4}}
\put(74.2,118.5){\circle{0.4}}
\put(75.1,120.){\circle{0.4}}
\put(76.,121.5){\circle{0.4}}
\put(76.8,123.){\circle{0.4}}
\put(77.5,124.5){\circle{0.4}}
\put(78.3,126.){\circle{0.4}}
\put(78.9,127.5){\circle{0.4}}
\put(79.6,129.){\circle{0.4}}
\put(80.1,130.5){\circle{0.4}}
\put(80.6,132.){\circle{0.4}}
\put(81.,133.5){\circle{0.4}}
\put(81.4,135.){\circle{0.4}}
\put(81.7,136.5){\circle{0.4}}
\put(82.,138.){\circle{0.4}}
\put(82.2,139.5){\circle{0.4}}
\put(82.3,141.){\circle{0.4}}
\put(82.3,142.5){\circle{0.4}}
\put(82.4,144.){\circle{0.4}}
\put(82.3,145.5){\circle{0.4}}
\put(82.2,147.){\circle{0.4}}
\put(82.1,148.5){\circle{0.4}}
\put(81.9,150.){\circle{0.4}}
\put(81.6,151.5){\circle{0.4}}
\put(81.4,153.){\circle{0.4}}
\put(81.,154.5){\circle{0.4}}
\put(80.7,156.){\circle{0.4}}
\put(80.3,157.5){\circle{0.4}}
\put(79.9,159.){\circle{0.4}}
\put(79.5,160.5){\circle{0.4}}
\put(79.1,162.){\circle{0.4}}
\put(78.6,163.5){\circle{0.4}}
\put(78.1,165.){\circle{0.4}}
\put(77.7,166.5){\circle{0.4}}
\put(77.2,168.){\circle{0.4}}
\put(76.7,169.5){\circle{0.4}}
\put(76.2,171.){\circle{0.4}}
\put(75.8,172.5){\circle{0.4}}
\put(75.3,174.){\circle{0.4}}
\put(74.9,175.5){\circle{0.4}}
\put(74.4,177.){\circle{0.4}}
\put(74.,178.5){\circle{0.4}}
\put(73.6,180.){\circle{0.4}}
\put(73.2,181.5){\circle{0.4}}
\put(72.9,183.){\circle{0.4}}
\put(72.6,184.5){\circle{0.4}}
\put(72.3,186.){\circle{0.4}}
\put(72.,187.5){\circle{0.4}}
\put(71.7,189.){\circle{0.4}}
\put(71.5,190.5){\circle{0.4}}
\put(71.3,192.){\circle{0.4}}
\put(71.1,193.5){\circle{0.4}}

\end{picture}

\newpage
\setlength{\unitlength}{1mm}
\setcounter{enumi}{-11}
\setcounter{enumii}{140}
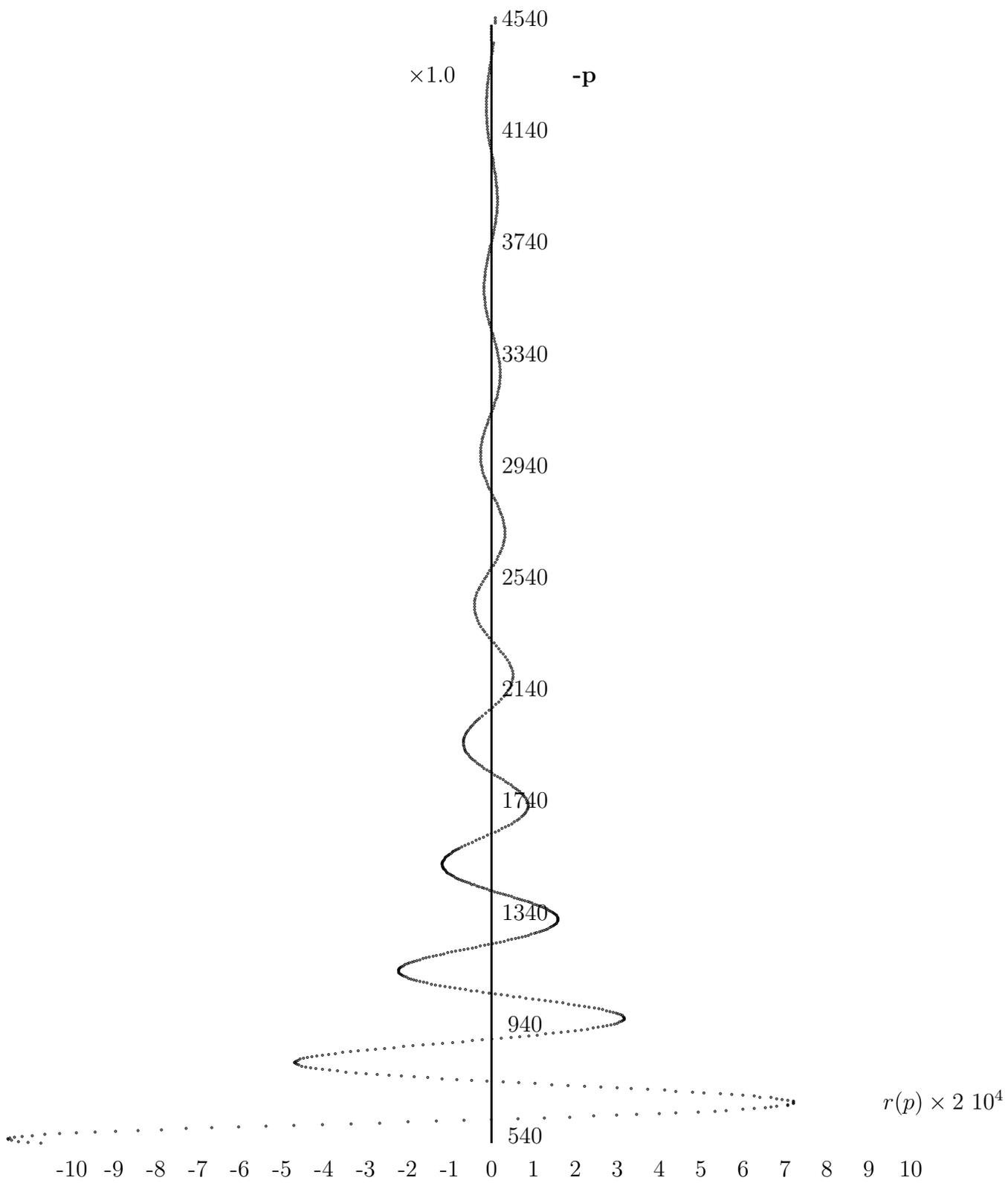
\begin{picture}(150,200)
\put(75,0){\line(0,1){200}}
\multiput(0,-6)(7.5,0){21}{\addtocounter{enumi}{1}\makebox(0,0)[b]
{\arabic{enumi}}}
\multiput(81,0)(0,20){11}{\addtocounter{enumii}{400}\makebox(0,0)[b]
{\arabic{enumii}}}
\put(145,6){$r(p) \times 2 \ {10}^{4}$}
\put(88,190){\bf{ -p} }
\put(60,190){$\times 1.0 $}
\put(-5.55,0){\circle{0.4}}
\put(-7.85,0.15){\circle{0.4}}
\put(-9.59,0.3){\circle{0.4}}
\put(-10.8,0.45){\circle{0.4}}
\put(-11.4,0.6){\circle{0.4}}
\put(-11.5,0.75){\circle{0.4}}
\put(-11.1,0.9){\circle{0.4}}
\put(-10.3,1.05){\circle{0.4}}
\put(-8.89,1.2){\circle{0.4}}
\put(-7.08,1.35){\circle{0.4}}
\put(-4.85,1.5){\circle{0.4}}
\put(-2.21,1.65){\circle{0.4}}
\put(0.794,1.8){\circle{0.4}}
\put(4.14,1.95){\circle{0.4}}
\put(7.81,2.1){\circle{0.4}}
\put(11.8,2.25){\circle{0.4}}
\put(16.,2.4){\circle{0.4}}
\put(20.4,2.55){\circle{0.4}}
\put(25.,2.7){\circle{0.4}}
\put(29.8,2.85){\circle{0.4}}
\put(34.7,3.){\circle{0.4}}
\put(39.7,3.15){\circle{0.4}}
\put(44.8,3.3){\circle{0.4}}
\put(49.9,3.45){\circle{0.4}}
\put(55.,3.6){\circle{0.4}}
\put(60.1,3.75){\circle{0.4}}
\put(65.2,3.9){\circle{0.4}}
\put(70.1,4.05){\circle{0.4}}
\put(75.,4.2){\circle{0.4}}
\put(79.8,4.35){\circle{0.4}}
\put(84.5,4.5){\circle{0.4}}
\put(89.,4.65){\circle{0.4}}
\put(93.3,4.8){\circle{0.4}}
\put(97.4,4.95){\circle{0.4}}
\put(101.,5.1){\circle{0.4}}
\put(105.,5.25){\circle{0.4}}
\put(108.,5.4){\circle{0.4}}
\put(112.,5.55){\circle{0.4}}
\put(115.,5.7){\circle{0.4}}
\put(117.,5.85){\circle{0.4}}
\put(120.,6.){\circle{0.4}}
\put(122.,6.15){\circle{0.4}}
\put(124.,6.3){\circle{0.4}}
\put(125.,6.45){\circle{0.4}}
\put(126.,6.6){\circle{0.4}}
\put(128.,6.75){\circle{0.4}}
\put(128.,6.9){\circle{0.4}}
\put(129.,7.05){\circle{0.4}}
\put(129.,7.2){\circle{0.4}}
\put(129.,7.35){\circle{0.4}}
\put(129.,7.5){\circle{0.4}}
\put(128.,7.65){\circle{0.4}}
\put(127.,7.8){\circle{0.4}}
\put(126.,7.95){\circle{0.4}}
\put(125.,8.1){\circle{0.4}}
\put(123.,8.25){\circle{0.4}}
\put(122.,8.4){\circle{0.4}}
\put(120.,8.55){\circle{0.4}}
\put(118.,8.7){\circle{0.4}}
\put(116.,8.85){\circle{0.4}}
\put(113.,9.){\circle{0.4}}
\put(111.,9.15){\circle{0.4}}
\put(108.,9.3){\circle{0.4}}
\put(106.,9.45){\circle{0.4}}
\put(103.,9.6){\circle{0.4}}
\put(100.,9.75){\circle{0.4}}
\put(97.5,9.9){\circle{0.4}}
\put(94.6,10.05){\circle{0.4}}
\put(91.7,10.2){\circle{0.4}}
\put(88.7,10.35){\circle{0.4}}
\put(85.8,10.5){\circle{0.4}}
\put(82.8,10.65){\circle{0.4}}
\put(79.9,10.8){\circle{0.4}}
\put(77.1,10.95){\circle{0.4}}
\put(74.2,11.1){\circle{0.4}}
\put(71.5,11.25){\circle{0.4}}
\put(68.8,11.4){\circle{0.4}}
\put(66.2,11.55){\circle{0.4}}
\put(63.6,11.7){\circle{0.4}}
\put(61.2,11.85){\circle{0.4}}
\put(58.9,12.){\circle{0.4}}
\put(56.7,12.15){\circle{0.4}}
\put(54.6,12.3){\circle{0.4}}
\put(52.6,12.45){\circle{0.4}}
\put(50.8,12.6){\circle{0.4}}
\put(49.,12.75){\circle{0.4}}
\put(47.5,12.9){\circle{0.4}}
\put(46.1,13.05){\circle{0.4}}
\put(44.8,13.2){\circle{0.4}}
\put(43.6,13.35){\circle{0.4}}
\put(42.7,13.5){\circle{0.4}}
\put(41.8,13.65){\circle{0.4}}
\put(41.1,13.8){\circle{0.4}}
\put(40.6,13.95){\circle{0.4}}
\put(40.2,14.1){\circle{0.4}}
\put(40.,14.25){\circle{0.4}}
\put(39.8,14.4){\circle{0.4}}
\put(39.9,14.55){\circle{0.4}}
\put(40.1,14.7){\circle{0.4}}
\put(40.4,14.85){\circle{0.4}}
\put(40.8,15.){\circle{0.4}}
\put(41.4,15.15){\circle{0.4}}
\put(42.1,15.3){\circle{0.4}}
\put(42.9,15.45){\circle{0.4}}
\put(43.8,15.6){\circle{0.4}}
\put(44.8,15.75){\circle{0.4}}
\put(45.9,15.9){\circle{0.4}}
\put(47.1,16.05){\circle{0.4}}
\put(48.4,16.2){\circle{0.4}}
\put(49.8,16.35){\circle{0.4}}
\put(51.2,16.5){\circle{0.4}}
\put(52.8,16.65){\circle{0.4}}
\put(54.3,16.8){\circle{0.4}}
\put(55.9,16.95){\circle{0.4}}
\put(57.6,17.1){\circle{0.4}}
\put(59.3,17.25){\circle{0.4}}
\put(61.,17.4){\circle{0.4}}
\put(62.8,17.55){\circle{0.4}}
\put(64.6,17.7){\circle{0.4}}
\put(66.3,17.85){\circle{0.4}}
\put(68.1,18.){\circle{0.4}}
\put(69.9,18.15){\circle{0.4}}
\put(71.6,18.3){\circle{0.4}}
\put(73.4,18.45){\circle{0.4}}
\put(75.1,18.6){\circle{0.4}}
\put(76.8,18.75){\circle{0.4}}
\put(78.5,18.9){\circle{0.4}}
\put(80.1,19.05){\circle{0.4}}
\put(81.7,19.2){\circle{0.4}}
\put(83.2,19.35){\circle{0.4}}
\put(84.7,19.5){\circle{0.4}}
\put(86.1,19.65){\circle{0.4}}
\put(87.5,19.8){\circle{0.4}}
\put(88.7,19.95){\circle{0.4}}
\put(90.,20.1){\circle{0.4}}
\put(91.1,20.25){\circle{0.4}}
\put(92.2,20.4){\circle{0.4}}
\put(93.2,20.55){\circle{0.4}}
\put(94.1,20.7){\circle{0.4}}
\put(95.,20.85){\circle{0.4}}
\put(95.7,21.){\circle{0.4}}
\put(96.4,21.15){\circle{0.4}}
\put(97.,21.3){\circle{0.4}}
\put(97.5,21.45){\circle{0.4}}
\put(97.9,21.6){\circle{0.4}}
\put(98.3,21.75){\circle{0.4}}
\put(98.5,21.9){\circle{0.4}}
\put(98.7,22.05){\circle{0.4}}
\put(98.8,22.2){\circle{0.4}}
\put(98.8,22.35){\circle{0.4}}
\put(98.7,22.5){\circle{0.4}}
\put(98.6,22.65){\circle{0.4}}
\put(98.4,22.8){\circle{0.4}}
\put(98.1,22.95){\circle{0.4}}
\put(97.7,23.1){\circle{0.4}}
\put(97.3,23.25){\circle{0.4}}
\put(96.8,23.4){\circle{0.4}}
\put(96.2,23.55){\circle{0.4}}
\put(95.6,23.7){\circle{0.4}}
\put(94.9,23.85){\circle{0.4}}
\put(94.2,24.){\circle{0.4}}
\put(93.4,24.15){\circle{0.4}}
\put(92.5,24.3){\circle{0.4}}
\put(91.7,24.45){\circle{0.4}}
\put(90.7,24.6){\circle{0.4}}
\put(89.8,24.75){\circle{0.4}}
\put(88.8,24.9){\circle{0.4}}
\put(87.8,25.05){\circle{0.4}}
\put(86.7,25.2){\circle{0.4}}
\put(85.6,25.35){\circle{0.4}}
\put(84.6,25.5){\circle{0.4}}
\put(83.5,25.65){\circle{0.4}}
\put(82.4,25.8){\circle{0.4}}
\put(81.2,25.95){\circle{0.4}}
\put(80.1,26.1){\circle{0.4}}
\put(79.,26.25){\circle{0.4}}
\put(77.9,26.4){\circle{0.4}}
\put(76.8,26.55){\circle{0.4}}
\put(75.7,26.7){\circle{0.4}}
\put(74.6,26.85){\circle{0.4}}
\put(73.5,27.){\circle{0.4}}
\put(72.4,27.15){\circle{0.4}}
\put(71.4,27.3){\circle{0.4}}
\put(70.4,27.45){\circle{0.4}}
\put(69.4,27.6){\circle{0.4}}
\put(68.5,27.75){\circle{0.4}}
\put(67.6,27.9){\circle{0.4}}
\put(66.7,28.05){\circle{0.4}}
\put(65.8,28.2){\circle{0.4}}
\put(65.,28.35){\circle{0.4}}
\put(64.3,28.5){\circle{0.4}}
\put(63.5,28.65){\circle{0.4}}
\put(62.9,28.8){\circle{0.4}}
\put(62.2,28.95){\circle{0.4}}
\put(61.6,29.1){\circle{0.4}}
\put(61.1,29.25){\circle{0.4}}
\put(60.6,29.4){\circle{0.4}}
\put(60.2,29.55){\circle{0.4}}
\put(59.8,29.7){\circle{0.4}}
\put(59.4,29.85){\circle{0.4}}
\put(59.1,30.){\circle{0.4}}
\put(58.9,30.15){\circle{0.4}}
\put(58.7,30.3){\circle{0.4}}
\put(58.5,30.45){\circle{0.4}}
\put(58.4,30.6){\circle{0.4}}
\put(58.4,30.75){\circle{0.4}}
\put(58.4,30.9){\circle{0.4}}
\put(58.4,31.05){\circle{0.4}}
\put(58.5,31.2){\circle{0.4}}
\put(58.6,31.35){\circle{0.4}}
\put(58.8,31.5){\circle{0.4}}
\put(59.,31.65){\circle{0.4}}
\put(59.3,31.8){\circle{0.4}}
\put(59.6,31.95){\circle{0.4}}
\put(60.,32.1){\circle{0.4}}
\put(60.3,32.25){\circle{0.4}}
\put(60.8,32.4){\circle{0.4}}
\put(61.2,32.55){\circle{0.4}}
\put(61.7,32.7){\circle{0.4}}
\put(62.2,32.85){\circle{0.4}}
\put(62.8,33.){\circle{0.4}}
\put(63.3,33.15){\circle{0.4}}
\put(63.9,33.3){\circle{0.4}}
\put(64.6,33.45){\circle{0.4}}
\put(65.2,33.6){\circle{0.4}}
\put(65.9,33.75){\circle{0.4}}
\put(66.5,33.9){\circle{0.4}}
\put(67.2,34.05){\circle{0.4}}
\put(67.9,34.2){\circle{0.4}}
\put(68.6,34.35){\circle{0.4}}
\put(69.4,34.5){\circle{0.4}}
\put(70.1,34.65){\circle{0.4}}
\put(70.8,34.8){\circle{0.4}}
\put(71.6,34.95){\circle{0.4}}
\put(72.3,35.1){\circle{0.4}}
\put(73.,35.25){\circle{0.4}}
\put(73.8,35.4){\circle{0.4}}
\put(74.5,35.55){\circle{0.4}}
\put(75.2,35.7){\circle{0.4}}
\put(75.9,35.85){\circle{0.4}}
\put(76.6,36.){\circle{0.4}}
\put(77.3,36.15){\circle{0.4}}
\put(78.,36.3){\circle{0.4}}
\put(78.6,36.45){\circle{0.4}}
\put(79.3,36.6){\circle{0.4}}
\put(79.9,36.75){\circle{0.4}}
\put(80.5,36.9){\circle{0.4}}
\put(81.1,37.05){\circle{0.4}}
\put(81.6,37.2){\circle{0.4}}
\put(82.1,37.35){\circle{0.4}}
\put(82.7,37.5){\circle{0.4}}
\put(83.1,37.65){\circle{0.4}}
\put(83.6,37.8){\circle{0.4}}
\put(84.,37.95){\circle{0.4}}
\put(84.4,38.1){\circle{0.4}}
\put(84.8,38.25){\circle{0.4}}
\put(85.1,38.4){\circle{0.4}}
\put(85.5,38.55){\circle{0.4}}
\put(85.7,38.7){\circle{0.4}}
\put(86.,38.85){\circle{0.4}}
\put(86.2,39.){\circle{0.4}}
\put(86.4,39.15){\circle{0.4}}
\put(86.6,39.3){\circle{0.4}}
\put(86.7,39.45){\circle{0.4}}
\put(86.8,39.6){\circle{0.4}}
\put(86.9,39.75){\circle{0.4}}
\put(86.9,39.9){\circle{0.4}}
\put(86.9,40.05){\circle{0.4}}
\put(86.9,40.2){\circle{0.4}}
\put(86.9,40.35){\circle{0.4}}
\put(86.8,40.5){\circle{0.4}}
\put(86.7,40.65){\circle{0.4}}
\put(86.6,40.8){\circle{0.4}}
\put(86.4,40.95){\circle{0.4}}
\put(86.2,41.1){\circle{0.4}}
\put(86.,41.25){\circle{0.4}}
\put(85.8,41.4){\circle{0.4}}
\put(85.5,41.55){\circle{0.4}}
\put(85.3,41.7){\circle{0.4}}
\put(85.,41.85){\circle{0.4}}
\put(84.6,42.){\circle{0.4}}
\put(84.3,42.15){\circle{0.4}}
\put(83.9,42.3){\circle{0.4}}
\put(83.6,42.45){\circle{0.4}}
\put(83.2,42.6){\circle{0.4}}
\put(82.8,42.75){\circle{0.4}}
\put(82.3,42.9){\circle{0.4}}
\put(81.9,43.05){\circle{0.4}}
\put(81.5,43.2){\circle{0.4}}
\put(81.,43.35){\circle{0.4}}
\put(80.5,43.5){\circle{0.4}}
\put(80.1,43.65){\circle{0.4}}
\put(79.6,43.8){\circle{0.4}}
\put(79.1,43.95){\circle{0.4}}
\put(78.6,44.1){\circle{0.4}}
\put(78.1,44.25){\circle{0.4}}
\put(77.6,44.4){\circle{0.4}}
\put(77.1,44.55){\circle{0.4}}
\put(76.6,44.7){\circle{0.4}}
\put(76.1,44.85){\circle{0.4}}
\put(75.6,45.){\circle{0.4}}
\put(75.1,45.15){\circle{0.4}}
\put(74.7,45.3){\circle{0.4}}
\put(74.2,45.45){\circle{0.4}}
\put(73.7,45.6){\circle{0.4}}
\put(73.3,45.75){\circle{0.4}}
\put(72.8,45.9){\circle{0.4}}
\put(72.4,46.05){\circle{0.4}}
\put(71.9,46.2){\circle{0.4}}
\put(71.5,46.35){\circle{0.4}}
\put(71.1,46.5){\circle{0.4}}
\put(70.7,46.65){\circle{0.4}}
\put(70.3,46.8){\circle{0.4}}
\put(69.9,46.95){\circle{0.4}}
\put(69.6,47.1){\circle{0.4}}
\put(69.2,47.25){\circle{0.4}}
\put(68.9,47.4){\circle{0.4}}
\put(68.6,47.55){\circle{0.4}}
\put(68.3,47.7){\circle{0.4}}
\put(68.,47.85){\circle{0.4}}
\put(67.8,48.){\circle{0.4}}
\put(67.6,48.15){\circle{0.4}}
\put(67.3,48.3){\circle{0.4}}
\put(67.1,48.45){\circle{0.4}}
\put(67.,48.6){\circle{0.4}}
\put(66.8,48.75){\circle{0.4}}
\put(66.7,48.9){\circle{0.4}}
\put(66.5,49.05){\circle{0.4}}
\put(66.4,49.2){\circle{0.4}}
\put(66.4,49.35){\circle{0.4}}
\put(66.3,49.5){\circle{0.4}}
\put(66.3,49.65){\circle{0.4}}
\put(66.2,49.8){\circle{0.4}}
\put(66.2,49.95){\circle{0.4}}
\put(66.2,50.1){\circle{0.4}}
\put(66.3,50.25){\circle{0.4}}
\put(66.3,50.4){\circle{0.4}}
\put(66.4,50.55){\circle{0.4}}
\put(66.5,50.7){\circle{0.4}}
\put(66.6,50.85){\circle{0.4}}
\put(66.7,51.){\circle{0.4}}
\put(66.8,51.15){\circle{0.4}}
\put(67.,51.3){\circle{0.4}}
\put(67.2,51.45){\circle{0.4}}
\put(67.4,51.6){\circle{0.4}}
\put(67.5,51.75){\circle{0.4}}
\put(67.8,51.9){\circle{0.4}}
\put(68.,52.05){\circle{0.4}}
\put(68.2,52.2){\circle{0.4}}
\put(68.5,52.35){\circle{0.4}}
\put(68.7,52.5){\circle{0.4}}
\put(69.,52.65){\circle{0.4}}
\put(69.3,52.8){\circle{0.4}}
\put(69.7,53.){\circle{0.4}}
\put(70.3,53.3){\circle{0.4}}
\put(70.9,53.6){\circle{0.4}}
\put(71.6,53.9){\circle{0.4}}
\put(72.3,54.2){\circle{0.4}}
\put(72.9,54.5){\circle{0.4}}
\put(73.6,54.8){\circle{0.4}}
\put(74.3,55.1){\circle{0.4}}
\put(75.,55.4){\circle{0.4}}
\put(75.7,55.7){\circle{0.4}}
\put(76.3,56.){\circle{0.4}}
\put(76.9,56.3){\circle{0.4}}
\put(77.5,56.6){\circle{0.4}}
\put(78.1,56.9){\circle{0.4}}
\put(78.7,57.2){\circle{0.4}}
\put(79.1,57.5){\circle{0.4}}
\put(79.6,57.8){\circle{0.4}}
\put(80.,58.1){\circle{0.4}}
\put(80.4,58.4){\circle{0.4}}
\put(80.7,58.7){\circle{0.4}}
\put(81.,59.){\circle{0.4}}
\put(81.2,59.3){\circle{0.4}}
\put(81.4,59.6){\circle{0.4}}
\put(81.5,59.9){\circle{0.4}}
\put(81.6,60.2){\circle{0.4}}
\put(81.6,60.5){\circle{0.4}}
\put(81.5,60.8){\circle{0.4}}
\put(81.5,61.1){\circle{0.4}}
\put(81.3,61.4){\circle{0.4}}
\put(81.2,61.7){\circle{0.4}}
\put(80.9,62.){\circle{0.4}}
\put(80.7,62.3){\circle{0.4}}
\put(80.4,62.6){\circle{0.4}}
\put(80.1,62.9){\circle{0.4}}
\put(79.7,63.2){\circle{0.4}}
\put(79.3,63.5){\circle{0.4}}
\put(78.9,63.8){\circle{0.4}}
\put(78.5,64.1){\circle{0.4}}
\put(78.,64.4){\circle{0.4}}
\put(77.6,64.7){\circle{0.4}}
\put(77.1,65.){\circle{0.4}}
\put(76.6,65.3){\circle{0.4}}
\put(76.1,65.6){\circle{0.4}}
\put(75.6,65.9){\circle{0.4}}
\put(75.1,66.2){\circle{0.4}}
\put(74.7,66.5){\circle{0.4}}
\put(74.2,66.8){\circle{0.4}}
\put(73.7,67.1){\circle{0.4}}
\put(73.3,67.4){\circle{0.4}}
\put(72.9,67.7){\circle{0.4}}
\put(72.5,68.){\circle{0.4}}
\put(72.1,68.3){\circle{0.4}}
\put(71.7,68.6){\circle{0.4}}
\put(71.4,68.9){\circle{0.4}}
\put(71.1,69.2){\circle{0.4}}
\put(70.9,69.5){\circle{0.4}}
\put(70.6,69.8){\circle{0.4}}
\put(70.4,70.1){\circle{0.4}}
\put(70.3,70.4){\circle{0.4}}
\put(70.2,70.7){\circle{0.4}}
\put(70.1,71.){\circle{0.4}}
\put(70.,71.3){\circle{0.4}}
\put(70.,71.6){\circle{0.4}}
\put(70.,71.9){\circle{0.4}}
\put(70.,72.2){\circle{0.4}}
\put(70.1,72.5){\circle{0.4}}
\put(70.2,72.8){\circle{0.4}}
\put(70.4,73.1){\circle{0.4}}
\put(70.5,73.4){\circle{0.4}}
\put(70.7,73.7){\circle{0.4}}
\put(70.9,74.){\circle{0.4}}
\put(71.2,74.3){\circle{0.4}}
\put(71.4,74.6){\circle{0.4}}
\put(71.7,74.9){\circle{0.4}}
\put(72.,75.2){\circle{0.4}}
\put(72.3,75.5){\circle{0.4}}
\put(72.6,75.8){\circle{0.4}}
\put(72.9,76.){\circle{0.4}}
\put(73.4,76.5){\circle{0.4}}
\put(74.,77.){\circle{0.4}}
\put(74.6,77.5){\circle{0.4}}
\put(75.2,78.){\circle{0.4}}
\put(75.8,78.5){\circle{0.4}}
\put(76.3,79.){\circle{0.4}}
\put(76.8,79.5){\circle{0.4}}
\put(77.3,80.){\circle{0.4}}
\put(77.7,80.5){\circle{0.4}}
\put(78.,81.){\circle{0.4}}
\put(78.3,81.5){\circle{0.4}}
\put(78.6,82.){\circle{0.4}}
\put(78.7,82.5){\circle{0.4}}
\put(78.8,83.){\circle{0.4}}
\put(78.9,83.5){\circle{0.4}}
\put(78.9,84.){\circle{0.4}}
\put(78.8,84.5){\circle{0.4}}
\put(78.6,85.){\circle{0.4}}
\put(78.4,85.5){\circle{0.4}}
\put(78.2,86.){\circle{0.4}}
\put(77.9,86.5){\circle{0.4}}
\put(77.5,87.){\circle{0.4}}
\put(77.1,87.5){\circle{0.4}}
\put(76.7,88.){\circle{0.4}}
\put(76.3,88.5){\circle{0.4}}
\put(75.9,89.){\circle{0.4}}
\put(75.5,89.5){\circle{0.4}}
\put(75.,90.){\circle{0.4}}
\put(74.6,90.5){\circle{0.4}}
\put(74.2,91.){\circle{0.4}}
\put(73.8,91.5){\circle{0.4}}
\put(73.5,92.){\circle{0.4}}
\put(73.1,92.5){\circle{0.4}}
\put(72.8,93.){\circle{0.4}}
\put(72.6,93.5){\circle{0.4}}
\put(72.4,94.){\circle{0.4}}
\put(72.2,94.5){\circle{0.4}}
\put(72.1,95.){\circle{0.4}}
\put(72.,95.5){\circle{0.4}}
\put(72.,96.){\circle{0.4}}
\put(72.,96.5){\circle{0.4}}
\put(72.,97.){\circle{0.4}}
\put(72.1,97.5){\circle{0.4}}
\put(72.3,98.){\circle{0.4}}
\put(72.4,98.5){\circle{0.4}}
\put(72.7,99.){\circle{0.4}}
\put(72.9,99.5){\circle{0.4}}
\put(73.2,100.){\circle{0.4}}
\put(73.5,100.5){\circle{0.4}}
\put(73.8,101.){\circle{0.4}}
\put(74.1,101.5){\circle{0.4}}
\put(74.4,102.){\circle{0.4}}
\put(74.7,102.5){\circle{0.4}}
\put(75.,103.){\circle{0.4}}
\put(75.4,103.5){\circle{0.4}}
\put(75.7,104.){\circle{0.4}}
\put(76.,104.5){\circle{0.4}}
\put(76.2,105.){\circle{0.4}}
\put(76.5,105.5){\circle{0.4}}
\put(76.7,106.){\circle{0.4}}
\put(76.9,106.5){\circle{0.4}}
\put(77.1,107.){\circle{0.4}}
\put(77.2,107.5){\circle{0.4}}
\put(77.3,108.){\circle{0.4}}
\put(77.4,108.5){\circle{0.4}}
\put(77.4,109.){\circle{0.4}}
\put(77.4,109.5){\circle{0.4}}
\put(77.4,110.){\circle{0.4}}
\put(77.3,110.5){\circle{0.4}}
\put(77.2,111.){\circle{0.4}}
\put(77.1,111.5){\circle{0.4}}
\put(77.,112.){\circle{0.4}}
\put(76.8,112.5){\circle{0.4}}
\put(76.6,113.){\circle{0.4}}
\put(76.4,113.5){\circle{0.4}}
\put(76.2,114.){\circle{0.4}}
\put(76.,114.5){\circle{0.4}}
\put(75.7,115.){\circle{0.4}}
\put(75.5,115.5){\circle{0.4}}
\put(75.2,116.){\circle{0.4}}
\put(75.,116.5){\circle{0.4}}
\put(74.7,117.){\circle{0.4}}
\put(74.5,117.5){\circle{0.4}}
\put(74.3,118.){\circle{0.4}}
\put(74.1,118.5){\circle{0.4}}
\put(73.9,119.){\circle{0.4}}
\put(73.7,119.5){\circle{0.4}}
\put(73.5,120.){\circle{0.4}}
\put(73.4,120.5){\circle{0.4}}
\put(73.3,121.){\circle{0.4}}
\put(73.2,121.5){\circle{0.4}}
\put(73.1,122.){\circle{0.4}}
\put(73.1,122.5){\circle{0.4}}
\put(73.1,123.){\circle{0.4}}
\put(73.1,123.5){\circle{0.4}}
\put(73.1,124.){\circle{0.4}}
\put(73.1,124.5){\circle{0.4}}
\put(73.2,125.){\circle{0.4}}
\put(73.3,125.5){\circle{0.4}}
\put(73.4,126.){\circle{0.4}}
\put(73.5,126.5){\circle{0.4}}
\put(73.7,127.){\circle{0.4}}
\put(73.8,127.5){\circle{0.4}}
\put(74.,128.){\circle{0.4}}
\put(74.2,128.5){\circle{0.4}}
\put(74.4,129.){\circle{0.4}}
\put(74.5,129.5){\circle{0.4}}
\put(74.7,130.){\circle{0.4}}
\put(74.9,130.5){\circle{0.4}}
\put(75.1,131.){\circle{0.4}}
\put(75.3,131.5){\circle{0.4}}
\put(75.5,132.){\circle{0.4}}
\put(75.6,132.5){\circle{0.4}}
\put(75.8,133.){\circle{0.4}}
\put(76.,133.5){\circle{0.4}}
\put(76.1,134.){\circle{0.4}}
\put(76.2,134.5){\circle{0.4}}
\put(76.3,135.){\circle{0.4}}
\put(76.4,135.5){\circle{0.4}}
\put(76.5,136.){\circle{0.4}}
\put(76.5,136.5){\circle{0.4}}
\put(76.6,137.){\circle{0.4}}
\put(76.6,137.5){\circle{0.4}}
\put(76.6,138.){\circle{0.4}}
\put(76.6,138.5){\circle{0.4}}
\put(76.5,139.){\circle{0.4}}
\put(76.5,139.5){\circle{0.4}}
\put(76.4,140.){\circle{0.4}}
\put(76.3,140.5){\circle{0.4}}
\put(76.2,141.){\circle{0.4}}
\put(76.1,141.5){\circle{0.4}}
\put(76.,142.){\circle{0.4}}
\put(75.9,142.5){\circle{0.4}}
\put(75.8,143.){\circle{0.4}}
\put(75.6,143.5){\circle{0.4}}
\put(75.5,144.){\circle{0.4}}
\put(75.3,144.5){\circle{0.4}}
\put(75.2,145.){\circle{0.4}}
\put(75.,145.5){\circle{0.4}}
\put(74.9,146.){\circle{0.4}}
\put(74.7,146.5){\circle{0.4}}
\put(74.6,147.){\circle{0.4}}
\put(74.5,147.5){\circle{0.4}}
\put(74.3,148.){\circle{0.4}}
\put(74.2,148.5){\circle{0.4}}
\put(74.1,149.){\circle{0.4}}
\put(74.,149.5){\circle{0.4}}
\put(73.9,150.){\circle{0.4}}
\put(73.9,150.5){\circle{0.4}}
\put(73.8,151.){\circle{0.4}}
\put(73.8,151.5){\circle{0.4}}
\put(73.7,152.){\circle{0.4}}
\put(73.7,152.5){\circle{0.4}}
\put(73.7,153.){\circle{0.4}}
\put(73.7,153.5){\circle{0.4}}
\put(73.7,154.){\circle{0.4}}
\put(73.8,154.5){\circle{0.4}}
\put(73.8,155.){\circle{0.4}}
\put(73.9,155.5){\circle{0.4}}
\put(73.9,156.){\circle{0.4}}
\put(74.,156.5){\circle{0.4}}
\put(74.1,157.){\circle{0.4}}
\put(74.2,157.5){\circle{0.4}}
\put(74.3,158.){\circle{0.4}}
\put(74.4,158.5){\circle{0.4}}
\put(74.5,159.){\circle{0.4}}
\put(74.6,159.5){\circle{0.4}}
\put(74.7,160.){\circle{0.4}}
\put(74.8,160.5){\circle{0.4}}
\put(75.,161.){\circle{0.4}}
\put(75.1,161.5){\circle{0.4}}
\put(75.2,162.){\circle{0.4}}
\put(75.3,162.5){\circle{0.4}}
\put(75.4,163.){\circle{0.4}}
\put(75.5,163.5){\circle{0.4}}
\put(75.6,164.){\circle{0.4}}
\put(75.7,164.5){\circle{0.4}}
\put(75.8,165.){\circle{0.4}}
\put(75.8,165.5){\circle{0.4}}
\put(75.9,166.){\circle{0.4}}
\put(76.,166.5){\circle{0.4}}
\put(76.,167.){\circle{0.4}}
\put(76.,167.5){\circle{0.4}}
\put(76.1,168.){\circle{0.4}}
\put(76.1,168.5){\circle{0.4}}
\put(76.1,169.){\circle{0.4}}
\put(76.1,169.5){\circle{0.4}}
\put(76.,170.){\circle{0.4}}
\put(76.,170.5){\circle{0.4}}
\put(76.,171.){\circle{0.4}}
\put(75.9,171.5){\circle{0.4}}
\put(75.9,172.){\circle{0.4}}
\put(75.8,172.5){\circle{0.4}}
\put(75.8,173.){\circle{0.4}}
\put(75.7,173.5){\circle{0.4}}
\put(75.6,174.){\circle{0.4}}
\put(75.5,174.5){\circle{0.4}}
\put(75.4,175.){\circle{0.4}}
\put(75.4,175.5){\circle{0.4}}
\put(75.3,176.){\circle{0.4}}
\put(75.2,176.5){\circle{0.4}}
\put(75.1,177.){\circle{0.4}}
\put(75.,177.5){\circle{0.4}}
\put(74.9,178.){\circle{0.4}}
\put(74.8,178.5){\circle{0.4}}
\put(74.7,179.){\circle{0.4}}
\put(74.6,179.5){\circle{0.4}}
\put(74.5,180.){\circle{0.4}}
\put(74.5,180.5){\circle{0.4}}
\put(74.4,181.){\circle{0.4}}
\put(74.3,181.5){\circle{0.4}}
\put(74.3,182.){\circle{0.4}}
\put(74.2,182.5){\circle{0.4}}
\put(74.2,183.){\circle{0.4}}
\put(74.2,183.5){\circle{0.4}}
\put(74.1,184.){\circle{0.4}}
\put(74.1,184.5){\circle{0.4}}
\put(74.1,185.){\circle{0.4}}
\put(74.1,185.5){\circle{0.4}}
\put(74.1,186.){\circle{0.4}}
\put(74.1,186.5){\circle{0.4}}
\put(74.1,187.){\circle{0.4}}
\put(74.2,187.5){\circle{0.4}}
\put(74.2,188.){\circle{0.4}}
\put(74.2,188.5){\circle{0.4}}
\put(74.3,189.){\circle{0.4}}
\put(74.3,189.5){\circle{0.4}}
\put(74.4,190.){\circle{0.4}}
\put(74.5,190.5){\circle{0.4}}
\put(74.5,191.){\circle{0.4}}
\put(74.6,191.5){\circle{0.4}}
\put(74.7,192.){\circle{0.4}}
\put(74.7,192.5){\circle{0.4}}
\put(74.8,193.){\circle{0.4}}
\put(74.9,193.5){\circle{0.4}}
\put(75.,194.){\circle{0.4}}
\put(75.,194.5){\circle{0.4}}
\put(75.1,195.){\circle{0.4}}
\put(75.2,195.5){\circle{0.4}}
\put(75.2,196.){\circle{0.4}}
\put(75.3,196.5){\circle{0.4}}
\put(75.4,197.){\circle{0.4}}
\put(75.,197.5){\circle{0.4}}
\put(75.,198.){\circle{0.4}}
\put(75.,198.5){\circle{0.4}}
\put(75.,199.){\circle{0.4}}
\put(75.,199.5){\circle{0.4}}
\put(75.,200.){\circle{0.4}}
\put(75.7,200.5){\circle{0.4}}
\put(75.7,201.){\circle{0.4}}
\put(75.7,201.5){\circle{0.4}}

\end{picture}
\newpage
\setlength{\unitlength}{1mm}
\setcounter{enumi}{-11}
\setcounter{enumii}{4820}
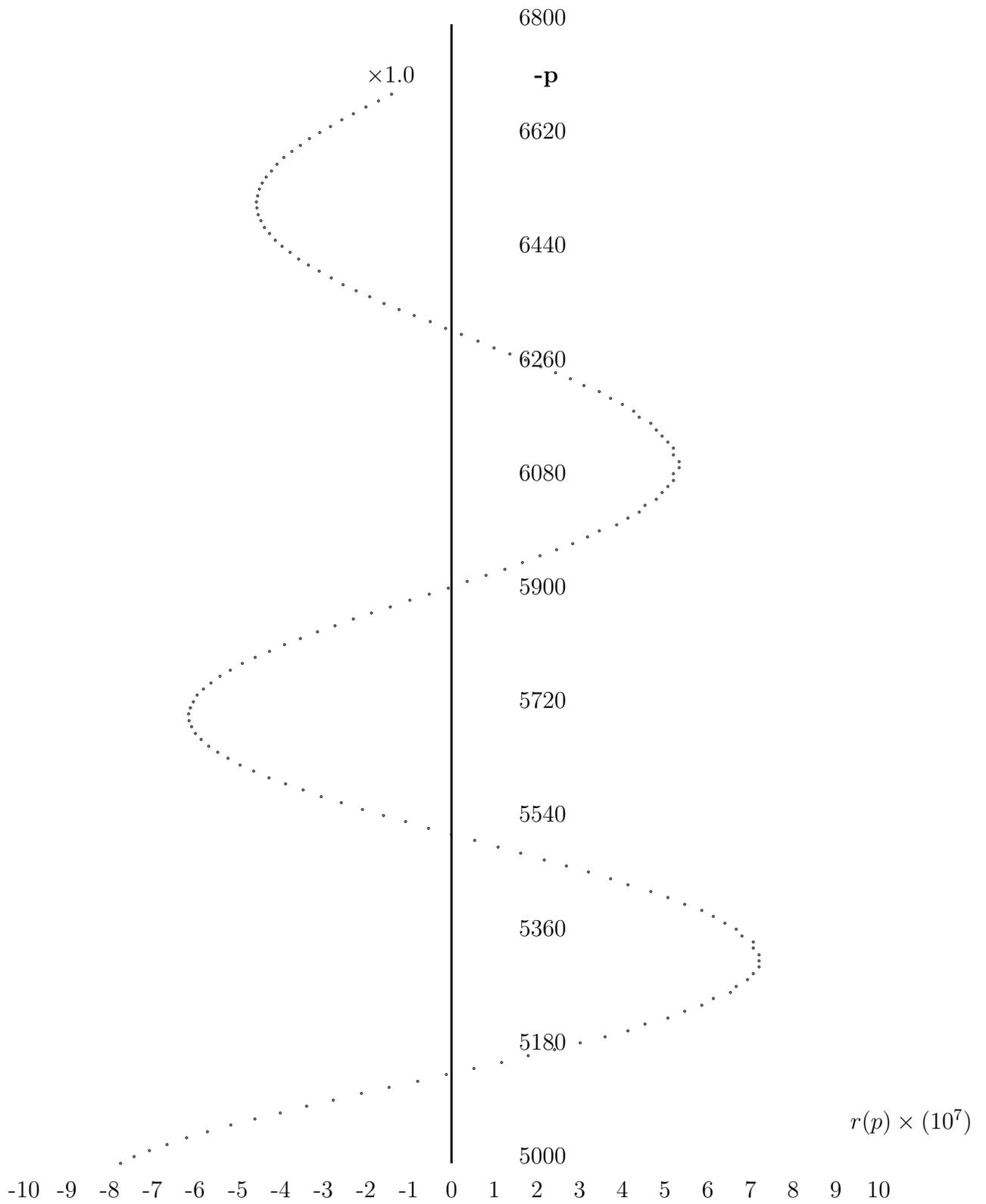
\begin{picture}(150,200)
\put(75,0){\line(0,1){200}}
\multiput(0,-6)(7.5,0){21}{\addtocounter{enumi}{1}\makebox(0,0)[b]
{\arabic{enumi}}}
\multiput(91,0)(0,20){11}{\addtocounter{enumii}{180}\makebox(0,0)[b]
{\arabic{enumii}}}
\put(145,6){$r(p) \times  ( {10}^{7})$}
\put(88,190){\bf{ -p} }
\put(60,190){$\times 1.0 $}
\put(16.9,0.){\circle{0.4}}
\put(19.3,1.111){\circle{0.4}}
\put(22.,2.222){\circle{0.4}}
\put(25.1,3.333){\circle{0.4}}
\put(28.5,4.444){\circle{0.4}}
\put(32.2,5.556){\circle{0.4}}
\put(36.3,6.667){\circle{0.4}}
\put(40.5,7.778){\circle{0.4}}
\put(45.,8.889){\circle{0.4}}
\put(49.6,10.){\circle{0.4}}
\put(54.3,11.11){\circle{0.4}}
\put(59.2,12.22){\circle{0.4}}
\put(64.1,13.33){\circle{0.4}}
\put(69.1,14.44){\circle{0.4}}
\put(74.1,15.56){\circle{0.4}}
\put(79.,16.67){\circle{0.4}}
\put(83.8,17.78){\circle{0.4}}
\put(88.5,18.89){\circle{0.4}}
\put(93.1,20.){\circle{0.4}}
\put(97.6,21.11){\circle{0.4}}
\put(102.,22.22){\circle{0.4}}
\put(106.,23.33){\circle{0.4}}
\put(109.,24.44){\circle{0.4}}
\put(113.,25.56){\circle{0.4}}
\put(116.,26.67){\circle{0.4}}
\put(119.,27.78){\circle{0.4}}
\put(121.,28.89){\circle{0.4}}
\put(124.,30.){\circle{0.4}}
\put(125.,31.11){\circle{0.4}}
\put(127.,32.22){\circle{0.4}}
\put(128.,33.33){\circle{0.4}}
\put(129.,34.44){\circle{0.4}}
\put(129.,35.56){\circle{0.4}}
\put(129.,36.67){\circle{0.4}}
\put(128.,37.78){\circle{0.4}}
\put(128.,38.89){\circle{0.4}}
\put(126.,40.){\circle{0.4}}
\put(125.,41.11){\circle{0.4}}
\put(123.,42.22){\circle{0.4}}
\put(121.,43.33){\circle{0.4}}
\put(119.,44.44){\circle{0.4}}
\put(116.,45.56){\circle{0.4}}
\put(113.,46.67){\circle{0.4}}
\put(110.,47.78){\circle{0.4}}
\put(106.,48.89){\circle{0.4}}
\put(103.,50.){\circle{0.4}}
\put(99.1,51.11){\circle{0.4}}
\put(95.2,52.22){\circle{0.4}}
\put(91.3,53.33){\circle{0.4}}
\put(87.2,54.44){\circle{0.4}}
\put(83.2,55.56){\circle{0.4}}
\put(79.1,56.67){\circle{0.4}}
\put(75.,57.78){\circle{0.4}}
\put(71.,58.89){\circle{0.4}}
\put(67.,60.){\circle{0.4}}
\put(63.1,61.11){\circle{0.4}}
\put(59.4,62.22){\circle{0.4}}
\put(55.7,63.33){\circle{0.4}}
\put(52.2,64.44){\circle{0.4}}
\put(49.,65.56){\circle{0.4}}
\put(45.9,66.67){\circle{0.4}}
\put(43.,67.78){\circle{0.4}}
\put(40.3,68.89){\circle{0.4}}
\put(37.9,70.){\circle{0.4}}
\put(35.8,71.11){\circle{0.4}}
\put(34.,72.22){\circle{0.4}}
\put(32.4,73.33){\circle{0.4}}
\put(31.1,74.44){\circle{0.4}}
\put(30.1,75.56){\circle{0.4}}
\put(29.4,76.67){\circle{0.4}}
\put(29.,77.78){\circle{0.4}}
\put(28.9,78.89){\circle{0.4}}
\put(29.2,80.){\circle{0.4}}
\put(29.7,81.11){\circle{0.4}}
\put(30.4,82.22){\circle{0.4}}
\put(31.5,83.33){\circle{0.4}}
\put(32.8,84.44){\circle{0.4}}
\put(34.4,85.56){\circle{0.4}}
\put(36.2,86.67){\circle{0.4}}
\put(38.3,87.78){\circle{0.4}}
\put(40.6,88.89){\circle{0.4}}
\put(43.1,90.){\circle{0.4}}
\put(45.7,91.11){\circle{0.4}}
\put(48.5,92.22){\circle{0.4}}
\put(51.5,93.33){\circle{0.4}}
\put(54.6,94.44){\circle{0.4}}
\put(57.8,95.56){\circle{0.4}}
\put(61.,96.67){\circle{0.4}}
\put(64.3,97.78){\circle{0.4}}
\put(67.7,98.89){\circle{0.4}}
\put(71.1,100.){\circle{0.4}}
\put(74.5,101.1){\circle{0.4}}
\put(77.8,102.2){\circle{0.4}}
\put(81.1,103.3){\circle{0.4}}
\put(84.4,104.4){\circle{0.4}}
\put(87.5,105.6){\circle{0.4}}
\put(90.6,106.7){\circle{0.4}}
\put(93.5,107.8){\circle{0.4}}
\put(96.3,108.9){\circle{0.4}}
\put(98.9,110.){\circle{0.4}}
\put(101.,111.1){\circle{0.4}}
\put(104.,112.2){\circle{0.4}}
\put(106.,113.3){\circle{0.4}}
\put(108.,114.4){\circle{0.4}}
\put(109.,115.6){\circle{0.4}}
\put(111.,116.7){\circle{0.4}}
\put(112.,117.8){\circle{0.4}}
\put(113.,118.9){\circle{0.4}}
\put(114.,120.){\circle{0.4}}
\put(114.,121.1){\circle{0.4}}
\put(115.,122.2){\circle{0.4}}
\put(115.,123.3){\circle{0.4}}
\put(114.,124.4){\circle{0.4}}
\put(114.,125.6){\circle{0.4}}
\put(113.,126.7){\circle{0.4}}
\put(112.,127.8){\circle{0.4}}
\put(111.,128.9){\circle{0.4}}
\put(110.,130.){\circle{0.4}}
\put(108.,131.1){\circle{0.4}}
\put(107.,132.2){\circle{0.4}}
\put(105.,133.3){\circle{0.4}}
\put(103.,134.4){\circle{0.4}}
\put(101.,135.6){\circle{0.4}}
\put(98.3,136.7){\circle{0.4}}
\put(95.9,137.8){\circle{0.4}}
\put(93.3,138.9){\circle{0.4}}
\put(90.7,140.){\circle{0.4}}
\put(88.,141.1){\circle{0.4}}
\put(85.3,142.2){\circle{0.4}}
\put(82.5,143.3){\circle{0.4}}
\put(79.7,144.4){\circle{0.4}}
\put(76.8,145.6){\circle{0.4}}
\put(74.,146.7){\circle{0.4}}
\put(71.3,147.8){\circle{0.4}}
\put(68.5,148.9){\circle{0.4}}
\put(65.8,150.){\circle{0.4}}
\put(63.2,151.1){\circle{0.4}}
\put(60.7,152.2){\circle{0.4}}
\put(58.3,153.3){\circle{0.4}}
\put(56.,154.4){\circle{0.4}}
\put(53.9,155.6){\circle{0.4}}
\put(51.8,156.7){\circle{0.4}}
\put(49.9,157.8){\circle{0.4}}
\put(48.2,158.9){\circle{0.4}}
\put(46.7,160.){\circle{0.4}}
\put(45.3,161.1){\circle{0.4}}
\put(44.1,162.2){\circle{0.4}}
\put(43.1,163.3){\circle{0.4}}
\put(42.2,164.4){\circle{0.4}}
\put(41.6,165.6){\circle{0.4}}
\put(41.1,166.7){\circle{0.4}}
\put(40.9,167.8){\circle{0.4}}
\put(40.8,168.9){\circle{0.4}}
\put(41.,170.){\circle{0.4}}
\put(41.3,171.1){\circle{0.4}}
\put(41.8,172.2){\circle{0.4}}
\put(42.5,173.3){\circle{0.4}}
\put(43.4,174.4){\circle{0.4}}
\put(44.4,175.6){\circle{0.4}}
\put(45.6,176.7){\circle{0.4}}
\put(47.,177.8){\circle{0.4}}
\put(48.5,178.9){\circle{0.4}}
\put(50.1,180.){\circle{0.4}}
\put(51.9,181.1){\circle{0.4}}
\put(53.8,182.2){\circle{0.4}}
\put(55.8,183.3){\circle{0.4}}
\put(57.8,184.4){\circle{0.4}}
\put(60.,185.6){\circle{0.4}}
\put(62.2,186.7){\circle{0.4}}
\put(64.5,187.8){\circle{0.4}}

\end{picture}
\newpage

\setlength{\unitlength}{1mm}
\setcounter{enumi}{-11}
\setcounter{enumii}{7130}
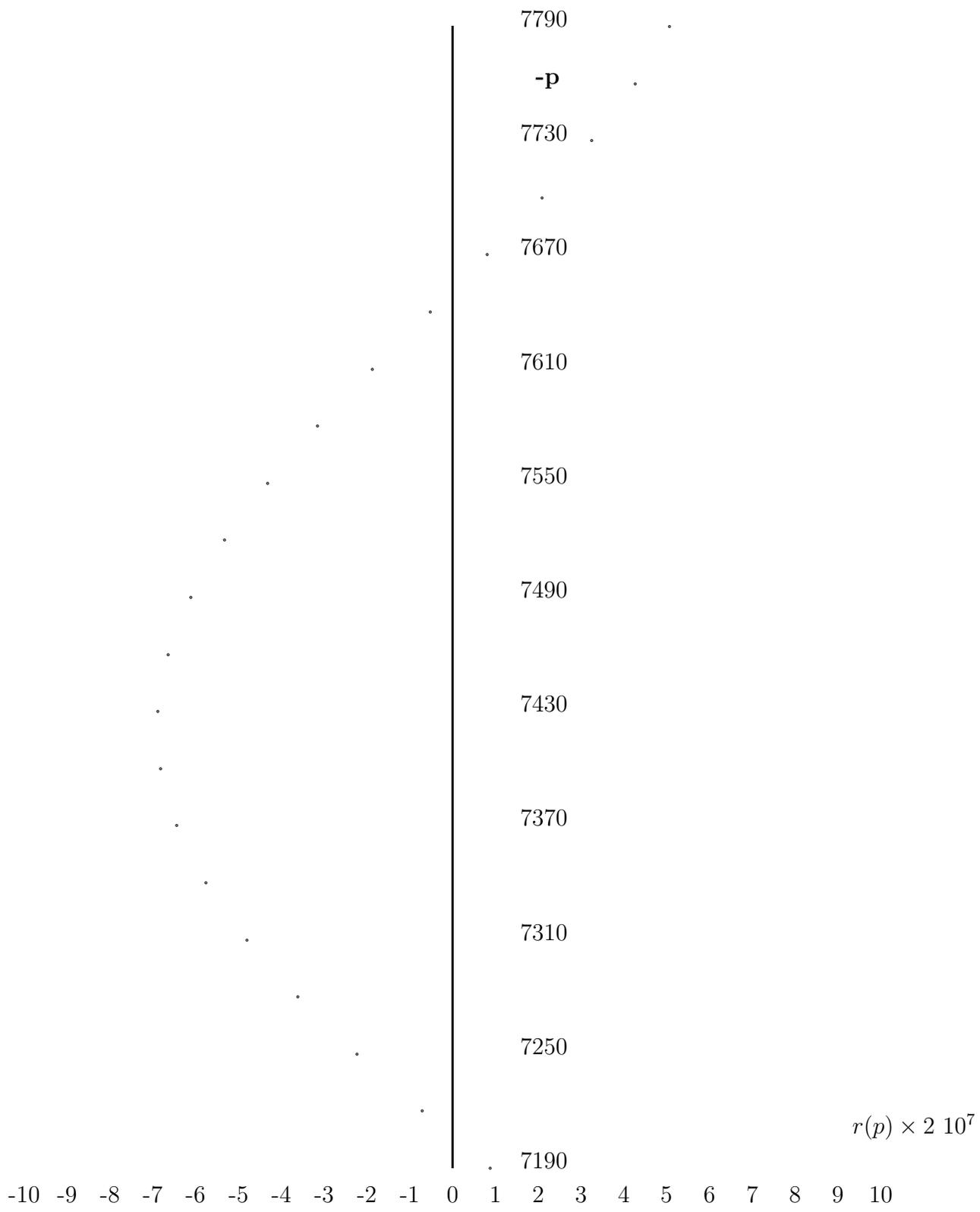
\begin{picture}(150,200)
\put(75,0){\line(0,1){200}}
\multiput(0,-6)(7.5,0){21}{\addtocounter{enumi}{1}\makebox(0,0)[b]
{\arabic{enumi}}}
\multiput(91,0)(0,20){11}{\addtocounter{enumii}{60}\makebox(0,0)[b]
{\arabic{enumii}}}
\put(145,6){$r(p) \times 2 \ {10}^{7}$}
\put(88,190){\bf{ -p} }
\put(81.6,0){\circle{0.4}}
\put(69.7,10.){\circle{0.4}}
\put(58.3,20.){\circle{0.4}}
\put(47.9,30.){\circle{0.4}}
\put(39.,40.){\circle{0.4}}
\put(31.8,50.){\circle{0.4}}
\put(26.7,60.){\circle{0.4}}
\put(23.9,70.){\circle{0.4}}
\put(23.4,80.){\circle{0.4}}
\put(25.2,90.){\circle{0.4}}
\put(29.2,100.){\circle{0.4}}
\put(35.1,110.){\circle{0.4}}
\put(42.6,120.){\circle{0.4}}
\put(51.4,130.){\circle{0.4}}
\put(61.,140.){\circle{0.4}}
\put(71.1,150.){\circle{0.4}}
\put(81.1,160.){\circle{0.4}}
\put(90.7,170.){\circle{0.4}}
\put(99.4,180.){\circle{0.4}}
\put(107.,190.){\circle{0.4}}
\put(113.,200.){\circle{0.4}}

\end{picture}

\subsection*{Acknowledgments}
Discussions with prof. Ooguri about the WKB formula in $QCD_3$ and large $N$ 
expansion is greatly appreciated. I am grateful to Dr. S. Cherkis for 
bringing to my attention the paper \cite{oog}, and  \cite{adsh},  
thus initiating this work,
and for discussions. I am grateful to M. Lapidus and I. Penkov and to 
for the hospitality during my  stay at UCR, and for conviniently locating the 
University of California, Riverside not far from the beach, 
Pasadena, and Santa Barbara, and to 
participants of the graduate seminar of prof. Lapidus for their interest to
this work. 

After completion of this work, the paper hepth-9806125 by 
R.de Mello Koch, A.Jevicki, M.Mihailescu, J.P.Nunes, which has some
overlap with our paper,  also appeared.

\end{document}